\documentclass[11pt]{article}

\usepackage[T1]{fontenc}
\usepackage[cp1251]{inputenc}
\usepackage{textcomp}
\usepackage[centertags]{amsmath}
\usepackage[mediummath]{nccmath}
\usepackage{amsfonts}
\usepackage{amssymb}
\usepackage{mathtools}
\usepackage[pdftex]{hyperref}
\usepackage{graphicx}
\usepackage[numbers,sort&compress]{natbib}

\usepackage{paperinitial}

\paperinitialization{15mm}{15mm}{15mm}{15mm}{2pt}{10pt}

\DeclareMathOperator{\im}{Im}

\DeclareMathOperator{\rot}{rot}

\newcommand{\lan}{\langle}
\newcommand{\ran}{\rangle}

\newcommand{\vf}{\varphi}
\newcommand{\vk}{\varkappa}
\newcommand{\s}{\sigma}

\newcommand{\al}{\alpha}
\newcommand{\be}{\beta}
\newcommand{\ga}{\gamma}
\newcommand{\Ga}{\Gamma}
\newcommand{\de}{\delta}
\newcommand{\De}{\Delta}
\newcommand{\ka}{\varkappa}
\newcommand{\la}{\lambda}

\newcommand{\spx}{\mathbf{x}}

\newcommand{\spp}{\mathbf{p}}
\newcommand{\spk}{\mathbf{k}}

\newcommand{\spe}{\mathbf{e}}
\newcommand{\spA}{\mathbf{A}}
\newcommand{\spR}{\mathbf{R}}
\newcommand{\spP}{\mathbf{P}}

\newcommand{\bs}{\boldsymbol}

\newcommand{\sixj}[6]{%
\begin{Bmatrix}
  #1 & #2 & #3 \\
  #4 & #5 & #6
\end{Bmatrix}}

\newcommand{\ninej}[9]{%
	\begin{Bmatrix}
		#1 & #2 & #3 \\
		#4 & #5 & #6 \\
		#7 & #8 & #9
\end{Bmatrix}}

\begin{document}
\allowdisplaybreaks[4]
\frenchspacing

\title{{\Large\textbf{Detection of twisted radiowaves with Rydberg atoms}}}

\date{}


\author{P.O. Kazinski\thanks{E-mail: \texttt{kpo@phys.tsu.ru}},\; P.S. Korolev\thanks{E-mail: \texttt{kizorph.d@gmail.com}},\;and V.A. Ryakin\thanks{E-mail: \texttt{vlad.r.a.phys@yandex.ru}}\\[0.5em]
{\normalsize  Physics Faculty, Tomsk State University, Tomsk 634050, Russia}\\[0.5em]
{\normalsize Mathematics and Mathematical Physics Division,}\\ {\normalsize Tomsk Polytechnic University, Tomsk 634050, Russia}
}

\maketitle

\begin{abstract}

The dynamics of the outer electron in an alkali atom in the presence of structured electromagnetic waves is described. The interaction of the alkali Rydberg atom with twisted radiowaves is considered. The two schemes for Rydberg-atom based detector of twisted radiowaves are proposed. According to the theoretical model for these detectors, they can record a source of twisted radiowaves with power down to several nW. The first scheme of the detector employs the nondipole transitions between Rydberg states induced by twisted radio photons. The second scheme involves the array of Rydberg-atom based antennas, every antenna measuring the dipole transitions excited by plane radiowaves comprising the twisted one.

\end{abstract}

\section{Introduction}

Recently, there has appeared a growing interest in using the vapors of alkali atoms for constructing highly sensitive detectors of electromagnetic waves in the radio frequency range working at room temperatures \cite{Sedlacek2012,Sedlacek2013,Anderson2014,Holloway2017,Simons2019,Song2019,Xue2021,Chen2025,Gong2024,Nowosielski2024,Hao2024,Ryabtsev2024,Yuan2023,Liu2022,Anderson2021AMFM,Robinson2021,Anderson2021,Stelmashenko2020,Jau2020,Gordon2019}. In these detectors, the outer electrons of alkali atoms are transferred by optical lasers to highly excited states -- the Rydberg states -- where these electrons are affected by radiowaves changing thereby the absorption coefficient for the optical probe laser. The change of the intensity of the probe laser is measured by a balanced photodetector. Despite simplicity of this scheme, such detectors possess an impressive sensitivity \cite{Gordon2019,Ryabtsev2024,Gong2024,Chen2025}. One of the reasons for that is the large magnitude of the matrix elements of the interaction Hamiltonian stemming from a large size of the region where the wave functions of the Rydberg states are different from zero.

Such peculiar properties of the Rydberg-atom based detectors can be used to record low-intensity twisted radiowaves \cite{Thide07,Mohammadi10,Roadmap16,SerboNew,ZWYB20,OAMPM,Noor22,JiangWerner22}. The twisted radiowaves or the radiowaves with nonzero orbital angular momentum are a promising tool for high-density information transfer in radio, terahertz, and optical ranges \cite{JiangWerner22,Roadmap16,SerboNew,Khan22,Willner22,ADKL20}. However, there is a principal impediment for a wide use of these waves in telecommunication. It is a considerable conical divergence of the intensity of twisted radiowaves at large distances from the source \cite{ZWYB20,JiangWerner22,Li20,Khan22,CZLZ23} that results in low intensity of such waves at the receiver. It is clear that the detector of twisted radiowaves with high sensitivity can mitigate this problem. In the present paper, we propose and describe theoretically the two schemes of the Rydberg-atom based receivers of twisted radiowaves that can record the signal with power down to several nW. In order to describe such detectors, we develop the theory for description of the electron dynamics in the alkali atoms irradiated by structured electromagnetic fields, in particular, by twisted photons. Notice that the various aspects of interaction of optical and hard twisted photons with atoms and atomic nuclei have been already investigated in the literature \cite{Afanas13,Mukherjee2018,Duan2019,Roadmap16,SerboNew,Lange2022,Peshkov2023,KazSokExcit,GiantResonance,Kirschbaum2024}. Furthermore, in the paper \cite{Wang2024}, the unsuccessful attempt to use the Rydberg-atom base detector to record the twisted radiowaves was reported. The complete theory for dynamics of the outer electron in an alkali atom in the structured external electromagnetic field was not constructed there and so the parameters of the system allowing to observe the effect of twisted radiowaves were not found.

The simplest realization of the Rydberg-atom based detector of plane radiowaves is based on a four level ladder system where the outer electron of an alkali atom evolves \cite{Sedlacek2012,Sedlacek2013,Holloway2017,Simons2019,Gordon2019,Jau2020,Song2019,Anderson2021,Stelmashenko2020,Robinson2021,Anderson2021AMFM,Hao2024,Ryabtsev2024,Chen2025}. The three lowest levels are attuned by the probe and coupling lasers transferring the electron between these levels in such a way that the electromagnetically induced transparency is realized for the probe laser \cite{Gea-Banacloche1995,Sandhya1997,McGloin2001,Cohen-Tannoudji_book_2004,Fleischhauer2005}. The absorption of the probe laser in the cuvette with vapor of alkali atoms is minimal in this case and increases drastically with a small change of the system parameters such as the detunings of frequencies of coherent sources of electromagnetic waves, the powers of these sources, and others. Therefore, if such a system is detuned from the electromagnetically induced transparency regime in some way, the intensity of the probe laser recorded by the photodetector changes appreciably. This property can be employed to record radiowaves that brings the outer electron from the third level to the adjacent Rydberg level and detunes thereby the three level system. The detectors of twisted radiowaves that we will consider are the modifications of this scheme. Their detailed descriptions are presented in Secs. \ref{Level_System}, \ref{Second_Scheme_Det}, and in Conclusion.

The paper is organized as follows. In Sec. \ref{General_Formulas}, we provide the general formulas describing interaction of the outer electron in an alkali atom with structured external electromagnetic field. In Sec. \ref{Hamiltonian}, we describe the model and the main approximations. Section \ref{Matr_Elem_H_Int} is devoted to evaluation of the matrix elements of the interaction Hamiltonian. In Sec. \ref{Detectors_Twisted_Radiowaves}, we describe the detectors of twisted radiowaves. In Sec. \ref{Bloch_Eqs}, we begin with the description of the dynamics of outer electron in an alkali atom. Then, in Sec. \ref{Dielectric_Suscept_Sec}, we obtain the explicit expression for the dielectric susceptibility of a gas of alkali atoms. Sections \ref{Level_System} and \ref{Second_Scheme_Det} are devoted to an extensive description of the Rydberg-atom based receivers of twisted radiowaves. In Conclusion we summarize the results. Some extensive calculations are removed to Appendices \ref{App_Mult_Trans} and \ref{App_Mult_Trans_Direct}, where we obtain the explicit expressions for the matrix elements of the interaction Hamiltonian. To this end, in Appendix \ref{App_Mult_Trans}, we employ the Siegert theorem to derive the matrix elements for $Ej$-transitions in the long-wavelength limit. In Appendix \ref{App_Mult_Trans_Direct}, we provide the direct evaluation of the matrix elements of the interaction Hamiltonian for both $Ej$- and $Mj$-transitions abandoning the long-wavelength approximation. We also deduce there the conditions ensuring that the matrix elements for $Ej$-transitions following from the Siegert theorem coincide with the matrix elements evaluated directly in the long-wavelength limit. Throughout the paper, we use the system of units such that $c=\hbar=1$ and $e^2=4\pi\al$, where $\alpha$ is the fine structure constant.

\section{General formulas}\label{General_Formulas}

\subsection{Hamiltonian}\label{Hamiltonian}

The Hamiltonian of an alkali atom with one active electron on the outer shell in the non-relativistic approximation reads
\begin{equation}
    \hat{H}=\hat{H}_0+\hat{H}_{int},
\end{equation}
where
\begin{equation}
    \hat{H}_0=\hat{H}_{cm}+\hat{H}_e.
\end{equation}
Here $\hat{H}_{cm}$ is the Hamiltonian of the atomic center of mass,
\begin{equation}\label{H_cm}
    \hat{H}_{cm}=\frac{\hat{\spP}^2}{2m_a},
\end{equation}
where $\spR$ is the vector of coordinates of the center of mass. For simplicity, let us assume that the cuvette containing the gas of alkali atoms is a rectangular box. The wave function of the center of mass satisfies the periodic boundary conditions at the boundaries of this box. The coordinates for the center of mass and relative motion are introduced as
\begin{equation}
    \spx_e=\spR+\frac{m_n}{m_a}\spx,\qquad \spx_n=\spR-\frac{m_e}{m_a}\spx,
\end{equation}
where $m_e$ is the electron mass, $m_n$ is the total mass of the nucleus of an atom and the electrons on closed shells (the mass of the atomic core), $m_a=m_e+m_n$ is the total mass of an atom, $\spx_e$ is the coordinate vector of the electron on the outer shell, $\spx_n$ is the coordinate vector of the center of mass of the atomic core. The momentum operator of the center of mass has the form $\hat{P}_i=-i\partial/\partial R_i$.

The Hamiltonian of the active electron takes the form
\begin{equation}\label{Hamilt_electron}
    \hat{H}_e=\frac{\hat{\spp}^2}{2\mu} +V_{mC}(\spx) +V_{so} +H_{hfs},
\end{equation}
where $\mu^{-1}=m_e^{-1}+m_n^{-1}$ is the reduced mass of the outer electron and the atomic core, $\hat{p}_i=-i\partial/\partial x_i$ is the operator of momentum of the electron motion relative to the center of mass, $V_{mC}(\spx)$ is the Coulomb potential modified due to the presence of electrons on the closed atomic shells \cite{Marinescu1994,Sibalic2017ARC}. We take into account the relativistic corrections to the Hamiltonian only in the form of spin-orbit interaction
\begin{equation}\label{V_SO}
    V_{so}=\frac{\al}{2 m^2_e r^3} (\mathbf{L}\mathbf{S}),\qquad r=|\spx|,
\end{equation}
where $\mathbf{L}$ is the orbital angular momentum of the outer electron, and $\mathbf{S}$ is its spin. The Hamiltonian of an hyperfine structure, $H_{hfs}$, has the standard form (see, e.g., \cite{Frish1963,LandLifshQM.11,Steck2024Cs,Steck_book_2025}). In the present paper, we will not take into account the hyperfine splitting due to the interaction $H_{hfs}$, since it is negligible for the Rydberg states (see, however, \cite{Ripka2022}). The states that take into account the hyperfine structure and the corresponding matrix elements of the interaction Hamiltonian are provided in Appendix \ref{App_Mult_Trans}.

The Hamiltonian of interaction with an external electromagnetic field can be cast into the form
\begin{equation}
    \hat{H}_{int}=\hat{H}_{int}^{(1)}+\hat{H}_{int}^{(2)}+\hat{H}_B,
\end{equation}
where
\begin{equation}\label{H_int12}
\begin{split}
    \hat{H}_{int}^{(1)}&=-\frac{e}{m_a}\spP\spA(t,\spR+\frac{m_n}{m_a}\spx)-e\spp \big[\frac{1}{m_e}\spA(t,\spR+\frac{m_n}{m_a}\spx) +\frac{1}{m_n}\spA(t,\spR-\frac{m_e}{m_a}\spx)\big],\\
    \hat{H}_{int}^{(2)}&=\frac{e^2}{2m_n}\spA^2(t,\spx_n)+\frac{e^2}{2m_e}\spA^2(t,\spx_e),
\end{split}
\end{equation}
where $\spA(t,\spx)$ is the potential of an external electromagnetic field in Coulomb gauge. It is assumed in Eqs. \eqref{H_int12} that the potential of the external electromagnetic field varies slowly on the scale of the atomic core, which is replaced by a point with coordinates $\spx_n$ in describing the interaction with the external electromagnetic field. The interaction of the magnetic moments of the electron and nucleus with an external electromagnetic field is written as
\begin{equation}\label{H_B}
    H_B=\mu_B\big[g_S\mathbf{S}\mathbf{H}(t,\spx_e) +g_I\mathbf{I} \mathbf{H}(t,\spx_n)\big],
\end{equation}
where $\mu_B=|e|/(2m_e)$ is the Bohr magneton, $\mathbf{H}=\rot\spA$, $\mathbf{S}$ is the electron spin operator, $\mathbf{I}$ is the orbital angular momentum operator of an atom, $g_S$ is the spin $g$-factor of an electron with anomalous magnetic moment taken into account, and $g_I$ is the $g$-factor of a nucleus. The spin $g$-factor of an electron is given by
\begin{equation}
    g_S\approx2+\al/\pi,
\end{equation}
where $\al$ is the fine structure constant. A more accurate value for this factor can be found in \cite{PDG2024}. The nuclear factor $g_I$ is determined by the structure of the nucleus and is taken from experiments. The numerical value of $g_I$ is small and the contribution of the corresponding term in \eqref{H_B} can be neglected as a rule. For example, $g_I\approx-3.99\times 10^{-4}$ for cesium, and $g_I\approx-9.95\times 10^{-4}$ for rubidium \cite{Steck2024Cs,Steck2024Rb}. Henceforth, we will not take this contribution into account. We also disregard the relativistic corrections to the interaction of the electromagnetic field with the electron arising from the $1/c$ expansion of the Dirac equation \cite{BjoDre}. Furthermore, we discard the contribution of $\hat{H}^{(2)}_{int}$. This contribution is of higher order than $\hat{H}^{(1)}_{int}$ with respect to the coupling constant and it is negligibly small with respect to the contribution of  $\hat{H}^{(1)}_{int}$ for the resonant interaction of an electron in an atom with an external electromagnetic wave when the frequency of the electromagnetic wave is close to the frequencies of the electron transitions between the atomic levels (see, e.g., \cite{Cohen-Tannoudji_book_2004}).

Let us suppose that the external electromagnetic field is in a coherent state with the complex amplitude $d_\la(\spk)$. Then the average value of electromagnetic potential operator becomes
\begin{equation}\label{electromag_potent}
    \spA(t,\spx)=\lan\hat{\spA}(\spx)\ran=\spA^{(+)}(t,\spx)+\spA^{(-)}(t,\spx),
\end{equation}
and the average number of photons in a given coherent state is written as
\begin{equation}
    N_\ga=\sum_\la\int d\spk |d_\la(\spk)|^2.
\end{equation}
Here
\begin{equation}
\begin{split}
     \spA^{(+)}(t,\spx)&=\sum_\la\int\frac{d\spk}{\sqrt{2(2\pi)^3k_0}} \spe_{(\la)}(\spk) e^{-ik_0t +i\spk\spx} d_\la(\spk) ,\qquad k_0=|\spk|,\\
     \spA^{(-)}(t,\spx)&=[\spA^{(+)}(t,\spx)]^*,
\end{split}
\end{equation}
and $\la=\pm1$ is the helicity of a photon with the polarization vector
\begin{equation}\label{photon_polarization}
\begin{split}
    \mathbf{e}_{(\la)}(\spk)=&\,(\cos\theta_k\cos\vf_k-i\la\sin\vf_k,\cos\theta_k\sin\vf_k+i\la\cos\phi_k,-\sin\theta_k)/\sqrt{2}=\\
    =&\,-\frac{\cos\theta_k+\la}{2}\spe_+ e^{-i\vf_k} +\frac{\cos\theta_k-\la}{2}\spe_- e^{i\vf_k}-\frac{\sin\theta_k}{\sqrt{2}}\spe_3.
\end{split}
\end{equation}
The definition of $\spe_{m}$ is given in Eq. \eqref{e_pm}, the angles $\theta_k$ and $\vf_k$ characterize the direction of the vector $\spk$, viz.,
\begin{equation}
    \spk=|\spk|(\sin\theta_k\cos\vf_k,\sin\theta_k\sin\vf_k,\cos\theta_k).
\end{equation}
Notice that there is a relation
\begin{equation}
    \mathbf{e}_{(\la)}(\spk)=-\la\spe_{\la},
\end{equation}
in the paraxial limit $\theta_k\rightarrow0$. Generally speaking, the Hamiltonian \eqref{Hamilt_electron} must also contain the terms describing the interaction of the electron with the quantum electromagnetic field $\hat{\spA}(\spx)$ and the Coulomb interaction of quantum electron fields (see, for example, \cite{KazLaz2021,KazMokRyk2025}). However, for our study, one does not need to take explicitly into account these contributions. The interaction with the quantum electromagnetic field, which describes, in particular, the emission and absorption of photons by an electron, will be effectively taken into account in the Bloch equations through the electron transition rates between states (see, e.g., \cite{Cohen-Tannoudji_book_2004,Happer2010}).

In order to describe the dynamics of the density matrix of an electron in an atom, it is necessary to find the matrix elements of the Hamiltonian of interaction of the classical electromagnetic field with the atom $\hat{H}^{(1)}_{int}+\hat{H}_B$. Let us simplify the expression for $\hat{H}^{(1)}_{int}$. The leading contribution to $\hat{H}^{(1)}_{int}$ comes from the first term in the square brackets (see \eqref{H_int12}). Let $r_B$ be a characteristic scale where the wave functions of the electron are concentrated for the states between which the transition is considered. If the transition occurs between the levels with significantly different $r_B$, then the maximum of the characteristic scales of these wave functions can be taken as $r_B$. The mention should be made that $r_B$ is of order $n^2a_0$, where $a_0$ is the Bohr radius and $n$ is the principal quantum number. The value of $r_B$ can be rather large for Rydberg states. In the case when
\begin{equation}\label{est1}
    k_0 r_B m_e/m_a\ll1,
\end{equation}
where $k_0$ is the energy of photons of the external electromagnetic wave, the first term in square brackets in $\hat{H}^{(1)}_{int}$ can be written as
\begin{equation}\label{H_orb}
    \hat{H}_{orb}=-e \frac{\spp}{m_e}\spA(t,\spR+\spx).
\end{equation}
The second term in the square brackets in the expression for $\hat{H}^{(1)}_{int}$ is suppressed in comparison with \eqref{H_orb} by the factor $m_e/m_n$. The first term in $\hat{H}^{(1)}_{int}$ is also small as compared with \eqref{H_orb}, since the rms velocity of an atom in a cell is much smaller than the rms velocity of an outer electron in an atom. If necessary, the contribution of the first term in $\hat{H}^{(1)}_{int}$ can be taken into account by perturbation theory.

When condition \eqref{est1} is met, it is also justified to replace $\spx_e\rightarrow \spR+\spx$ in the interaction Hamiltonian of the electron spin with magnetic field \eqref{H_B}. As a result, the Hamiltonian that we will use to describe the interaction of the electromagnetic field with the active electron in the alkali atom becomes
\begin{equation}\label{H_int_app}
    \hat{H}_{int}^{app}[\spA(t,\spR+\spx)]=-e \frac{\spp}{m_e}\spA(t,\spR+\spx)+\mu_B g_S\mathbf{S}\mathbf{H}(t,\spR+\spx).
\end{equation}

\subsection{Matrix elements of the interaction Hamiltonian}\label{Matr_Elem_H_Int}

\subsubsection{Separation of the center of mass}\label{Matr_Elem_H_Int_CM}

The complete set of eigenfunctions of the center-of-mass Hamiltonian \eqref{H_cm} is given by
\begin{equation}\label{wave_func_CM}
    f_\spP(\spR)=\frac{1}{\sqrt{V}}e^{i\spP\spR},
\end{equation}
where $V$ is the volume of a cuvette, and the momentum $\spP$ has a quasi-continuous spectrum of values such that the functions \eqref{wave_func_CM} satisfy the periodic boundary conditions at the boundary of the cell. We need to determine the matrix elements of the operator \eqref{H_int_app} in the basis of states
\begin{equation}
    |\spP,nLJM\ran=f_\spP(\spR)\psi_{nLJM}(\spx),
\end{equation}
where $\psi_{nLJM}(\spx)$ are the eigenstates of the operator $\hat{H}_e$, $n$ is the principal quantum number, and
\begin{equation}
\begin{aligned}
    \mathbf{J}^2\psi_{nLJM}&=J(J+1)\psi_{nLJM},&\qquad J_z\psi_{nLJM}&=M\psi_{nLJM},\\
    \qquad \mathbf{L}^2\psi_{nLJM}&=L(L+1)\psi_{nLJM},&\qquad \mathbf{S}^2\psi_{nLJM}&=\frac34\psi_{nLJM},
\end{aligned}
\end{equation}
where $\mathbf{J}=\mathbf{L}+\mathbf{S}$. Let us separate the dependence of the matrix element on the dynamics of the center of mass of the atom
\begin{equation}\label{matr_elem_ham_int}
    \lan \spP',n'L'J'M'|\hat{H}_{int}^{app}[\spA(t,\spR+\spx)]|\spP,nLJM\ran=\int \frac{d\spR}{V} e^{i(\spP-\spP')\spR} \mathcal{A}_{n'L'J'M',nLJM}(\spR),
\end{equation}
where
\begin{equation}
    \mathcal{A}_{n'L'J'M';nLJM}(\spR):=\lan n'L'J'M'|\hat{H}_{int}^{app}[\spA(t,\spR+\spx)]|nLJM\ran.
\end{equation}
Substituting the explicit form of the electromagnetic field potential \eqref{electromag_potent} into the expression for $\mathcal{A}_{n'L'J'M',nLJM}$, we obtain
\begin{equation}
    \mathcal{A}_{n'L'J'M';nLJM}(\spR)= \mathcal{A}^{(+)}_{n'L'J'M';nLJM}(\spR) + \mathcal{A}^{(-)}_{n'L'J'M';nLJM}(\spR),
\end{equation}
where
\begin{equation}\label{A_+_R}
    \mathcal{A}^{(+)}_{n'L'J'M';nLJM}(\spR)=\sum_\la \int\frac{d\spk d_\la(\spk)}{\sqrt{2(2\pi)^3k_0}} e^{-ik_0t+i\spk\spR} \lan n'L'J'M'|\hat{H}_{int}^{app}[\spe_{(\la)}(\spk) e^{i\spk\spx}]|nLJM\ran,
\end{equation}
and
\begin{equation}
    \mathcal{A}^{(-)}_{n'L'J'M';nLJM}(\spR)=\mathcal{A}^{(+)\dag}_{n'L'J'M';nLJM}(\spR).
\end{equation}
Employing the multipole expansion of the plane wave \eqref{mult_expans}, we come to
\begin{equation}\label{A_matr_elem}
\begin{split}
    \mathcal{A}^{(+)}_{n'L'J'M';nLJM}(\spR)=&\sum_{j=1}^\infty\sum_{m=-j}^j\sum_{p,\la} (i\la)^pi^j\sqrt{2\pi(2j+1)} \int\frac{d\spk d_\la(\spk)}{\sqrt{2(2\pi)^3k_0}} e^{-ik_0t+i\spk\spR}
    D^j_{m\la}(\vf_k,\theta_k,0)\times\\
    &\times \lan n'L'J'M'|\hat{H}_{int}^{app}[\bs\psi^p_{jm}(k_0,\spx)]|nLJM\ran.
\end{split}
\end{equation}
Applying the Wigner-Eckart theorem, the matrix element on the last line can be conveniently written in terms of the reduced matrix element
\begin{equation}\label{H_int_matr_elem}
    \lan n'L'J'M'|\hat{H}_{int}^{app}[\bs\psi^p_{jm}(k_0,\spx)]|nLJM\ran=\frac{1}{\sqrt{2J'+1}}C^{J'M'}_{JMjm}
    \lan n'L'J'\|\hat{H}_{int}^{app}[\bs\psi^p_{j}(k_0)]\|nLJ\ran,
\end{equation}
where the dependence on $M$, $M'$, and $m$ is explicitly singled out in the form of the Clebsch–Gordan coefficient $C^{J'M'}_{JMjm}$. It specifies, in particular, the standard selection rules
\begin{equation}\label{sel_rule_ang_mom}
    |J'-J|\leqslant j\leqslant J'+J,\qquad M'=M+m,\qquad |M'-M|\leqslant j.
\end{equation}
The states of the outer electron in an alkali atom have parity $(-1)^L$. Therefore, the parity selection rule:
\begin{equation}\label{sel_rule_parity}
    L+L'+j+1+p \text{ is even}.
\end{equation}
The matrix elements \eqref{H_int_matr_elem} are calculated in Appendices \ref{App_Mult_Trans} and \ref{App_Mult_Trans_Direct}. In particular, in the long-wavelength limit the matrix elements of $Ej$-transitions have the form \eqref{reduced_matr_elem} and the matrix elements of $Mj$-transitions in the same approximation are given in formulas \eqref{Mj_longwavelength_i}, \eqref{Mj_longwavelength_ii}.

If the external electromagnetic field has the form of a plane wave, viz., $d_\la(\spk)$ is concentrated near a particular $\spk_\ga$, the matrix element \eqref{A_matr_elem} can be cast into the form
\begin{equation}
    \mathcal{A}^{(+)}_{n'L'J'M';nLJM}(\spR)=e^{i\spk_\ga\spR}\mathcal{A}^{(+)}_{n'L'J'M';nLJM}(0),
\end{equation}
where the matrix element $\mathcal{A}^{(+)}_{n'L'J'M';nLJM}(0)$ does not depend on $\spR$. Then the matrix element of the interaction Hamiltonian \eqref{matr_elem_ham_int} is written as
\begin{equation}
\begin{split}
    \lan \spP',n'L'J'M'|\hat{H}_{int}^{app}[\spA(t,\spR+\spx)]|\spP,nLJM\ran=&\,\frac{(2\pi)^3}{V}\de(\spP'-\spP-\spk_\ga) \mathcal{A}^{(+)}_{n'L'J'M';nLJM}(0)+\\
    &+\frac{(2\pi)^3}{V}\de(\spP'-\spP+\spk_\ga) \mathcal{A}^{(-)}_{n'L'J'M';nLJM}(0).
\end{split}
\end{equation}
In the present paper, in investigating the dynamics of an outer electron of an alkali atom, we will neglect the change of the momentum of atomic center of mass due to influence of the laser plane-wave radiation on it. So we will assume that
\begin{equation}
    \frac{(2\pi)^3}{V}\de(\spP'-\spP\pm\spk_\ga)\approx\frac{(2\pi)^3}{V}\de(\spP'-\spP)=\de_{\spP',\spP},
\end{equation}
and, consequently,
\begin{equation}
    \lan \spP',n'L'J'M'|\hat{H}_{int}^{app}[\spA(t,\spR+\spx)]|\spP,nLJM\ran\approx \de_{\spP',\spP}\mathcal{A}_{n'L'J'M';nLJM}(0).
\end{equation}
In the case where the external electromagnetic field does not have a form of a plane wave, we will assume that the estimate is valid,
\begin{equation}\label{estim_twisted}
    k_0 L_{cuv}\ll1,
\end{equation}
where $L_{cuv}$ is the typical size of a cuvette with the alkali atoms. Such a field for the system we consider has the form of a twisted radiowave. Then
\begin{equation}
    \mathcal{A}_{n'L'J'M';nLJM}(\spR)\approx \mathcal{A}_{n'L'J'M';nLJM}(0),
\end{equation}
and
\begin{equation}\label{matr_elem_ham_int_app}
    \lan \spP',n'L'J'M'|\hat{H}_{int}^{app}[\spA(t,\spR+\spx)]|\spP,nLJM\ran=\de_{\spP',\spP} \mathcal{A}_{n'L'J'M';nLJM}(0).
\end{equation}
Thus, in both cases within the specified approximations, the matrix element of the interaction Hamiltonian has the form \eqref{matr_elem_ham_int_app}.

The plane-wave and twisted external electromagnetic fields can be obtained by setting
\begin{equation}\label{cohrent state_ampl}
    d_\la(\spk)=e^{im_\ga\vf_k} a_\la(k_\perp,k_z),\qquad a_\la(k_\perp,k_z)= a(k_\perp,k_z) \de_{\la s},
\end{equation}
where $m_\ga$ is the projection of the total angular momentum of each photon in a coherent state, $k_\perp=|k_x+ik_y|$, and $s$ is the helicity of the photons. In the paraxial regime with $m_\ga=s$ and $k_\perp\ll k_0$,  the complex amplitude \eqref{cohrent state_ampl} describes the coherent state of photons with a plane wavefront and circular polarization $s$. Otherwise, $d_\la(\spk)$ corresponds to twisted radiowaves. We will further assume that
\begin{equation}\label{e-m_field}
    d_\la(\spk)=\sum_{r=1}^{N_L}e^{i m_\ga^r\vf_k}a^r_\la(k_\perp,k_z),\qquad a^r_\la(k_\perp,k_z)=a^r(k_\perp,k_z) \de_{\la s_r},
\end{equation}
where $a^r(k_\perp,k_z)$ is concentrated near the values of photon energies $k_0^r$, and $N_L$ is the number of sources of coherent radiation with different frequencies $k_0^r$, helicities $s_r$, and projections of the total angular momentum $m_\ga^r$. Then going to a cylindrical coordinate system with respect to $\spk$ in the integral $\mathcal{A}^{(+)}_{n'L'J'M';nLJM}(0)$ (see \eqref{A_matr_elem}) and performing the integral with respect to the azimuth angle $\vf_k$ with account for the explicit form \eqref{Wigner_matr} of the Wigner matrices, we approximately obtain
\begin{equation}\label{matr_elem_3}
     \lan \spP',n'L'J'M'|\hat{H}_{int}^{app}[\spA(t,\spR+\spx)]|\spP,nLJM\ran\approx\de_{\spP',\spP} \sum_{r=1}^{N_L} \big(V^r_{n'L'J'M';nLJM} e^{-ik_0^rt} +V^{r\dag}_{n'L'J'M';nLJM} e^{ik_0^rt} \big),
\end{equation}
where
\begin{equation}\label{V_matr_elem}
    V^r_{n'L'J'M';nLJM}=\sum_{j=1}^\infty\sum_{p=0,1}  i^j\sqrt{\frac{2j+1}{2J'+1}} C^{J'M'}_{JMjm_\ga^r} h^{pj}_{m_\ga^r} \lan n'L'J'\|\hat{H}_{int}^{app}[\bs\psi^p_{j}(k^r_0)]\|nLJ\ran,
\end{equation}
and
\begin{equation}\label{h_j_m}
    h^{pj}_{m_\ga^r}:=\sum_\la (i\la)^p \int\frac{dk_z dk_\perp k_\perp}{\sqrt{2k_0}} d^j_{m_\ga^r\la}(\theta_k) a^r_\la(k_\perp,k_z)= (is_r)^p \int\frac{dk_z dk_\perp k_\perp}{\sqrt{2k_0}} d^j_{m_\ga^rs_r}(\theta_k) a^r(k_\perp,k_z).
\end{equation}
Here, the factors $\exp(\pm ik_0t)$ have been taken out of the integral over $\spk$. This is valid only if the linewidth of the radiation source $\De k_0 t\ll1$, where $t$ is the lapse of time where the evolution of the system is considered. In the case we are interested in, this lapse of time has to be much greater than the lifetimes of the active electron states where the dynamics described by the Bloch equations develop. This imposes rigorous restrictions on the linewidth. The effect of a non-zero linewidth of the radiation source can be effectively taken into account in the Bloch equations by adding the corresponding corrections to the energy of photons in the electromagnetic wave \cite{Gea-Banacloche1995,Downes2023}.

\subsubsection{Rotation of the coordinate system}

\begin{figure}[tp]
\centering
\includegraphics*[width=0.8\linewidth]{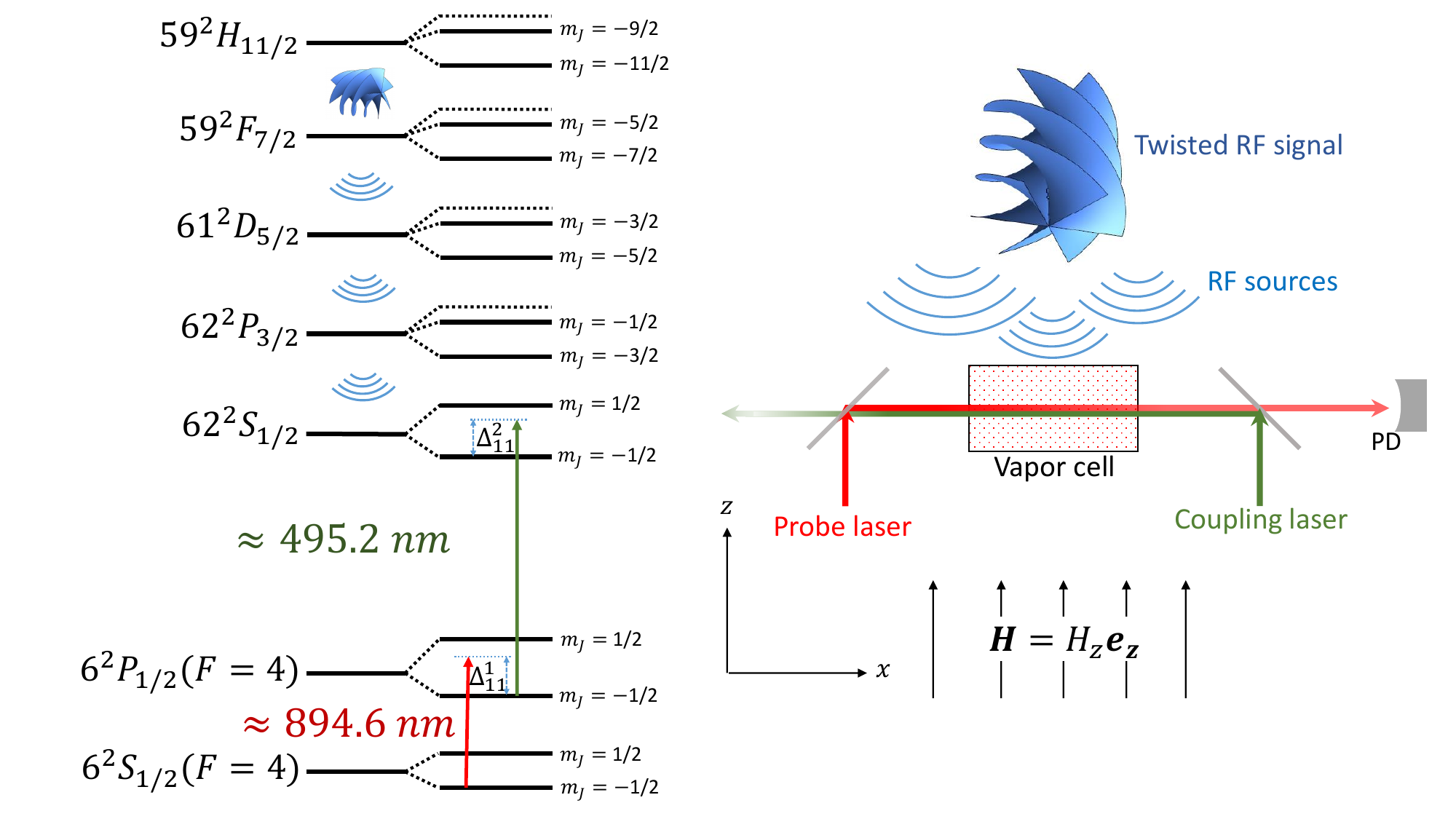}
\caption{{\footnotesize The scheme of the Rydberg-atom based detector of twisted radiowaves described in Sec. \ref{Level_System}. The signal of twisted radiowaves is heterodyned.}}
\label{scheme_7level_plots}
\end{figure}

In Eq. \eqref{V_matr_elem}, it is assumed that the axis of propagation of the electromagnetic wave interacting with the atom coincides with the quantization axis of the angular momentum projection. In our case, it is convenient to choose this axis (the $z$ axis) to coincide with the quantization axis of the total angular momentum for the twisted radiowave. This axis also coincides with the direction of the external magnetic field acting on the cuvette with the gas of alkali atoms (see Fig. \ref{scheme_7level_plots}). In that case, expression \eqref{V_matr_elem} cannot be used to describe the interaction of an electron in an atom with a plane laser wave propagating at an angle to the $z$ axis. To rotate the propagation axis, we can use the general formula \eqref{rotation_Wigner} by applying it to the multipoles in the multipole expansion \eqref{mult_expans}. Then it is easy to see that expression \eqref{V_matr_elem} is modified as
\begin{equation}\label{V_matr_elem_rotated}
    V^r_{n'L'J'M';nLJM}=\sum_{j=1}^\infty\sum_{p=0,1}\sum_{m=-j}^j i^j\sqrt{\frac{2j+1}{2J'+1}} C^{J'M'}_{JMjm} D^j_{mm_\ga^r}(\al_r,\be_r,\ga_r) h^{pj}_{m_\ga^r} \lan n'L'J'\|\hat{H}_{int}^{app}[\bs\psi^p_{j}(k^r_0)]\|nLJ\ran,
\end{equation}
where $\al_r$, $\be_r$, and $\ga_r$ are the Euler angles of rotation that transform the $z$ axis into the axis along which the $r$-th laser wave propagates, and the angle $\theta_k$ in expression \eqref{h_j_m} is measured from the axis of wave propagation. Of course, it is possible to remove the summation over $m$ in expression \eqref{V_matr_elem_rotated} by putting $m=M'-M$.

For a plane electromagnetic wave, one ought to set $d^j_{m_\ga^rs_r}(\theta_k)=\de_{m^r_\ga s_r}$ in \eqref{h_j_m}, i.e., the projection of the total angular momentum, $m_\ga^r$, is equal to the helicity of this wave $s_r$. The matrix element \eqref{V_matr_elem_rotated} for an electromagnetic wave with polarization distinct from a circular one is obtained by the corresponding linear combination of the matrix elements \eqref{V_matr_elem_rotated}. Notice that even in the case of a plane electromagnetic wave the transitions of any multipolarity $j$ are excited. Since $k_0 r_B\ll1$ in the long-wavelength limit and the matrix element of the $2^j$-th multipole transition is proportional to $(k_0 r_B)^j$, the main contribution comes from the transition with the lowest multipolarity admitted by the selection rules.

In the dipole approximation, we obviously have
\begin{equation}\label{V_matr_elem_rotated_dip}
    V^r_{n'L'J'M';nLJM}=i\sum_{p=0,1}\sum_{m=-1}^1 \sqrt{\frac{3}{2J'+1}} C^{J'M'}_{JM1m} D^1_{mm_\ga^r}(\al_r,\be_r,\ga_r) h^{p1}_{m_\ga^r} \lan n'L'J'\|\hat{H}_{int}^{app}[\bs\psi^p_{1}(k^r_0)]\|nLJ\ran.
\end{equation}
The matrix element \eqref{V_matr_elem_rotated_dip} is proportional to the matrix element of the operators of the electric or magnetic dipole moments of the atom (see, e.g., \cite{LandLifQED}) contracted with the polarization vector of the incident electromagnetic radiation (for the electric transitions) or with the vector product of the wave vector and the polarization vector (for the magnetic transitions). Therefore, if the incident electromagnetic wave is linearly polarized along the $z$ axis, then electric dipole transitions are possible only with $M'=M$. If the radiation is polarized in the plane orthogonal to the $z$ axis, then electric dipole transitions are possible only with $|M'-M|=1$. For magnetic dipole transitions, the situation is the opposite: the radiation linearly polarized along the $z$ axis induces the transitions with $|M'-M|=1$, whereas the radiation polarized in the plane orthogonal to the $z$ axis gives rise to the transitions with $M'=M$. The other selection rules for dipole radiation are the special cases of the general selection rules \eqref{sel_rule_ang_mom}, \eqref{sel_rule_parity} with $j=1$.

\subsubsection{Description of the external electromagnetic field}

In order to obtain $h^{pj}_{m_\ga^r}$, it is necessary to specify the profiles of wave functions of photons interacting with atoms. Let us take the Laguerre-Gaussian wave functions as such profiles
\begin{equation}\label{Laguerre_Gauss}
    a(k_\perp,k_z)=\sqrt{\frac{n_\ga!}{2\pi(n_\ga+|l|)!}} \sqrt{\frac{N_\ga}{\sqrt{2\pi}\s_z\s^2_\perp}} \Big(\frac{k_\perp}{\sqrt{2}\s_\perp}\Big)^{|l|} L_{n_\ga}^{|l|}\Big(\frac{k_\perp^2}{2\s_\perp^2}\Big) e^{-\frac{k_\perp^2}{4\s_\perp^2}} e^{-\frac{(k_z-k_z^0)^2}{4\s_z^2}},
\end{equation}
where we do not write the index $r$ for the wave function parameters to avoid cluttering with the notation, $l:=m_\ga-s$, $\s_\perp$, and $\s_z$ are the transverse and longitudinal rms deviations of momenta in the photon wave function, $k_z^0$ is the average value of the longitudinal component of the photon momentum. We will assume that $k_z^0\gg\s_z$ and $k_z^0\gg|l|^{1/2}\s_\perp$, i.e., we will consider the paraxial approximation. In this case, $\theta_k$ entering into $d^j_{m_\ga s}(\theta_k)$ is small and we can employ the asymptotic expression (4.18.3) of \cite{Varshalovich}:
\begin{equation}
    d^j_{m_\ga s}(\theta_k)\approx C^j_{m_\ga s} \theta_k^{|l|}\approx C^j_{m_\ga s}\big(\frac{k_\perp}{k_0}\big)^{|l|},
\end{equation}
where
\begin{equation}
    C^j_{m_\ga s} =\frac{(-1)^{(l+|l|)/2}}{2^{|l|}|l|!}\bigg[\frac{(j+\frac{|l|+|m_\ga+s|}{2})! (j+\frac{|l|-|m_\ga+s|}{2})!}{(j-\frac{|l|+|m_\ga+s|}{2})! (j-\frac{|l|-|m_\ga+s|}{2})!} \bigg]^{1/2}.
\end{equation}
Then, under the above conditions on $\s_z$ and $\s_\perp$, one can take approximately $k_0\approx k_z^0\approx k_0^r$ in the integral \eqref{h_j_m}. The integral over $k_z$ in \eqref{h_j_m} becomes Gaussian and the integral over $k_\perp$ reduces to (2.19.3.3) of \cite{Prud2}:
\begin{equation}
    \frac12\int_0^\infty dx x^{|l|} L_{n_\ga}^{|l|}(x)e^{-x/2}=(-1)^{n_\ga} 2^{|l|} \frac{(n_\ga+|l|)!}{n_\ga!}.
\end{equation}
As a result, we obtain
\begin{equation}\label{h_pjm_app}
    h^{pj}_{m_\ga}=(-1)^{n_\ga} (is)^p \sqrt{\frac{(n_\ga+|l|)!}{n_\ga!}} C^j_{m_\ga s}\Big(\frac{2\sqrt{2}\s_\perp}{k_0}\Big)^{|l|+1} \sqrt{\frac{k_0 N_\ga \s_z}{2(2\pi)^{1/2}}}.
\end{equation}
As long as the duration of the external electromagnetic field pulse is of order $1/\s_z$, the quantity
\begin{equation}\label{power_expl}
    P=k_0 N_\ga \s_z
\end{equation}
characterizes the power of the radiation source at a given frequency. For a plane electromagnetic wave $l=0$, $n_\ga=0$, and
\begin{equation}
    h^{pj}_{s}=(is)^p \frac{2\s_\perp}{k_0} \sqrt{\frac{k_0 N_\ga \s_z}{(2\pi)^{1/2}}}=:(is)^p h_{pw},
\end{equation}
where $m_\ga=s$.

To amplify the received signal on the photodetector recording the intensity of the probe laser passed through the cell with alkali atoms, a scheme with heterodyne is used \cite{Simons2019,Gordon2019,Robinson2021,Yuan2023,Gong2024}. Namely, along with the electromagnetic wave carrying information, the atom is irradiated by a reference electromagnetic wave with approximately the same frequency but with a greater amplitude (the local oscillator). As a result, the atom is affected by the interference of these waves that, as we will see, leads to an increase in the sensitivity of the detector we investigate. We suppose that the local oscillator obeys all the conditions specified above in deriving the explicit expressions for the matrix elements of the interaction Hamiltonian and $h^{pj}_{m_\ga}$. Moreover, we will assume that the local oscillator and the information wave have the same circular polarization $s$ and projections of the total angular momentum $m_\ga$. In this case, we have
\begin{equation}
    a(k_\perp,k_z)=a_0(k_\perp,k_z)+A_{sig} a_{sig}(k_\perp,k_z),
\end{equation}
in the expression for the external electromagnetic field \eqref{e-m_field}, where $a_0(k_\perp,k_z)$ describes the reference electromagnetic wave, $A_{sig} a_{sig}(k_\perp,k_z)$ corresponds to the electromagnetic wave carrying information, and $A_{sig}$ specifies the complex amplitude of the information electromagnetic wave relative to the reference wave. This amplitude does not depend on $\spk$ and is approximately constant at times of the order of and less than the inverse frequency of the electromagnetic wave at issue. Then only the expression for $h^{pj}_{m_\ga}$ changes in the matrix element of the interaction Hamiltonian \eqref{V_matr_elem}.

To derive the explicit expression for $h^{pj}_{m_\ga}$, we take the functions $a_0(k_\perp,k_z)$ and $a_{sig}(k_\perp,k_z)$ to be the Laguerre-Gaussian modes \eqref{Laguerre_Gauss} with their own $n_\ga$, $\s_\perp$, and $k_z^0\approx k_0$ in the paraxial regime. We denote these parameters as $n_0$, $\s^0_\perp$, and $k_0^0$ for for $a_0(k_\perp,k_z)$ and as $n_{sig}$, $\s^{sig}_\perp$, and $k_0^{sig}$ for $a_{sig}(k_\perp,k_z)$. The energies of photons $k_0^0$ and $k_0^{sig}$ are supposed to be close to each other. The remaining parameters in $a_0(k_\perp,k_z)$ and $a_{sig}(k_\perp,k_z)$ are identical. The power of the information wave $P_{sig}$ turned on separately from the reference wave is related to the power of the reference wave $P_0$ turned on separately from the information wave as
\begin{equation}
    P_{sig}=P_0 \frac{k_0^{sig}}{k_0^0}|A_{sig}|^2.
\end{equation}
Repeating the calculations given above in deriving the expression for $h^{pj}_{m_\ga}$, we obtain
\begin{equation}
    h^{pj}_{m_\ga}= (-1)^{n_0}(is)^p\sqrt{\frac{(n_0+|l|)!}{n_0!}} C^j_{m_\ga s}\Big(\frac{2\sqrt{2}\s_\perp}{k_0}\Big)^{|l|+1} \sqrt{\frac{P_0}{2(2\pi)^{1/2}}} \Big(1+\varkappa\sqrt{\frac{k_0^{sig}}{k_0^0}} A_{sig}\Big),
\end{equation}
where
\begin{equation}
    \varkappa:=(-1)^{n_{sig}-n_0} \sqrt{\frac{(n_{sig}+|l|)!n_0!}{(n_{0}+|l|)!n_{sig}!}} \Big(\frac{\s_\perp^{sig} k_0^0}{\s_\perp^0 k_0^{sig}}\Big)^{|l|+1}.
\end{equation}
The expression for $h^{pj}_{m_\ga}$ can be cast into the form \eqref{h_pjm_app} written in terms of the power \eqref{power_expl}
\begin{equation}\label{h_pjm_app_heter}
    h^{pj}_{m_\ga}=(-1)^{n_0}(is)^p\sqrt{\frac{(n_0+|l|)!}{n_0!}} C^j_{m_\ga s}\Big(\frac{2\sqrt{2}\s_\perp}{k_0}\Big)^{|l|+1} \sqrt{\frac{P_{eff}}{2(2\pi)^{1/2}}},
\end{equation}
where an effective complex power has been introduced
\begin{equation}\label{P_eff_expl}
    P_{eff}:=P_0 \Big(1+\varkappa\sqrt{\frac{k_0^{sig}}{k_0^0}} A_{sig}\Big)^2.
\end{equation}
In other words, with the above approximations, the account for heterodyning is reduced to replacing the signal power with the effective power \eqref{P_eff_expl}. So long as the power of the information wave is much smaller than the power of the local oscillator, one can suppose
\begin{equation}
    \big|\varkappa \sqrt{k_0^{sig}/k_0^0} A_{sig}\big|\ll1.
\end{equation}
Consequently, we obtain
\begin{equation}\label{P_eff_modul}
    |P_{eff}|\approx P_0+ 2\varkappa \sqrt{P_0P_{sig}} \cos(\arg A_{sig})
\end{equation}
for the modulus of effective power, i.e., the heterodyne amplifies the information signal. As for the plane electromagnetic wave, one has to set $n_0=n_{sig}=0$ and $l=0$ in the above expressions.

\subsubsection{Matrix elements for $Ej$-transitions}

For the level system discussed further in Sec. \ref{Level_System}, the main contribution to the matrix elements of the interaction Hamiltonian is given by $Ej$-transitions. Taking expression \eqref{reduced_matr_elem} for the reduced matrix elements of $Ej$-transitions in the long-wavelength limit, we have
\begin{equation}\label{E1_long_wavelngth}
\begin{split}
    V_{n'L'J'M';nLJM}=&\,ies(-1)^{1/2+J+L'+1} h_{pw} C^{J'M'}_{JM1s} \sqrt{\frac{(2J+1)(2L+1)}{2\pi}} \sixj{J'}{J}{1}{L}{L'}{1/2}\times\\ &\times C^{L'0}_{L010} E_{n'L'J'M';nLJM} r_{n'L'J';nLJ},
\end{split}
\end{equation}
for $E1$-transition induced by a plane wave with circular polarization $s$ propagating along the $z$ axis. As far as $Ej$-transitions induced by a twisted photon with total angular momentum projection $m_\ga$ and spin $s$ propagating along the $z$ axis are concerned, we obtain
\begin{equation}\label{Ej_long_wavelngth}
\begin{split}
    V_{n'L'J'M';nLJM}=&\,i^{j-1}e(-1)^{1/2+J+L'+j} h^{1j}_{m_\ga} C^{J'M'}_{JMjm_\ga} \sqrt{\frac{j+1}{j}} \sqrt{\frac{(2J+1)(2L+1)}{4\pi}} \sixj{J'}{J}{j}{L}{L'}{1/2}\times\\
    &\times C^{L'0}_{L0j0} \frac{E_{n'L'J'M';nLJM}k_0^{j-1}}{(2j-1)!!} r^j_{n'L'J';nLJ},
\end{split}
\end{equation}
in the same approximation. The matrix elements of the operator $r^j$ can be estimated as $r^j_B$, where $r_B$ was defined near formula \eqref{est1}. Then it is evident that, for the same source powers, the matrix elements of the interaction Hamiltonian are suppressed for transitions induced by twisted photons in comparison with the analogous matrix elements for transitions induced by plane-wave photons by a factor of
\begin{equation}
    \Big(\frac{2\sqrt{2}\s_\perp}{k_0}\Big)^{|l|} (k_0 r_B)^{j-1}.
\end{equation}
This estimate shows, in particular, that in order to enhance the influence of twisted electromagnetic waves on an electron in an alkali atom, $\s_\perp$ of the source of twisted photons should not be too small as compared with $k_0$.

For the detector scheme we consider in the present paper, the probe and coupling lasers propagate along the $x$ axis (see Fig. \ref{scheme_7level_plots}) and are described by a plane-wave electromagnetic field. In order to obtain the corresponding matrix elements of the interaction Hamiltonian in the dipole approximation, we employ expression \eqref{V_matr_elem_rotated_dip}, where $\al_r=\ga_r=0$ and $\be_r=\pi/2$. Such a transformation rotates the $z$ axis into the $x$ axis, while the $y$ axis remains unchanged. Inasmuch as \cite{Varshalovich}
\begin{equation}
\begin{split}
    d^1_{mm'}\big(\frac{\pi}{2}\big)=&\,\frac12(\de_{m1}\de_{m'1} +\de_{m,-1}\de_{m',-1} +\de_{m1}\de_{m',-1} +\de_{m,-1}\de_{m'1})+\\ &+\frac1{\sqrt{2}} (\de_{m0}\de_{m'1} -\de_{m1}\de_{m'0} +\de_{m,-1}\de_{m'0} -\de_{m0}\de_{m',-1}),
\end{split}
\end{equation}
we come to
\begin{equation}\label{V_matr_elem_rotated_dip_E}
\begin{split}
    V^r_{n'L'J'M';nLJM}|_{m^r_\ga=1}&=-h^r_{pw} \sqrt{\frac{3}{2J'+1}} \lan n'L'J'\|\hat{H}_{int}^{app}[\bs\psi^1_{1}(k^r_0)]\|nLJ\ran \big(\frac12 C^{J'M'}_{JM11} +\frac1{\sqrt{2}}C^{J'M'}_{JM10} +\frac12C^{J'M'}_{JM1,-1} \big),\\
    V^r_{n'L'J'M';nLJM}|_{m^r_\ga=-1}&=-h^r_{pw} \sqrt{\frac{3}{2J'+1}}  \lan n'L'J'\|\hat{H}_{int}^{app}[\bs\psi^1_{1}(k^r_0)]\|nLJ\ran \big(-\frac12 C^{J'M'}_{JM11} +\frac1{\sqrt{2}}C^{J'M'}_{JM10} -\frac12C^{J'M'}_{JM1,-1} \big),
\end{split}
\end{equation}
for $E1$-transitions with $m^r_\ga=s^r=\pm1$. For comparison, let us give similar expressions for $M1$-transitions
\begin{equation}\label{V_matr_elem_rotated_dip_M}
\begin{split}
    V^r_{n'L'J'M';nLJM}|_{m^r_\ga=1}&=ih^r_{pw} \sqrt{\frac{3}{2J'+1}} \lan n'L'J'\|\hat{H}_{int}^{app}[\bs\psi^0_{1}(k^r_0)]\|nLJ\ran \big(\frac12 C^{J'M'}_{JM11} +\frac1{\sqrt{2}}C^{J'M'}_{JM10} +\frac12C^{J'M'}_{JM1,-1} \big),\\
    V^r_{n'L'J'M';nLJM}|_{m^r_\ga=-1}&=ih^r_{pw} \sqrt{\frac{3}{2J'+1}}  \lan n'L'J'\|\hat{H}_{int}^{app}[\bs\psi^0_{1}(k^r_0)]\|nLJ\ran \big(\frac12 C^{J'M'}_{JM11} -\frac1{\sqrt{2}}C^{J'M'}_{JM10} +\frac12C^{J'M'}_{JM1,-1} \big).
\end{split}
\end{equation}
To deduce the matrix element for $E1$-transition induced by a plane laser wave polarized along the $z$ axis, we add expressions \eqref{V_matr_elem_rotated_dip_E} and divide by $-\sqrt{2}$. Then we obtain
\begin{equation}\label{V_matr_elem_rotated_dip_E_ez}
    V^r_{n'L'J'M';nLJM}=h^r_{pw} \sqrt{\frac{3}{2J'+1}} C^{J'M'}_{JM10}\lan n'L'J'\|\hat{H}_{int}^{app}[\bs\psi^1_{1}(k^r_0)]\|nLJ\ran .
\end{equation}
The extra minus sign in this expression arises from the fact that, in rotating around the $y$ axis, the unit vector $\spe_x$ goes to $-\spe_z$. It is seen that the selection rule $M'=M$ discussed after Eq. \eqref{V_matr_elem_rotated_dip} is fulfilled. The other selection rules for the magnetic quantum numbers discussed there readily follow from expressions \eqref{V_matr_elem_rotated_dip_E} and \eqref{V_matr_elem_rotated_dip_M}.

The explicit expression for the matrix element \eqref{V_matr_elem_rotated_dip_E_ez} in the long-wavelength limit is obtained by substituting \eqref{reduced_matr_elem}:
\begin{equation}\label{V_dip_expl}
\begin{split}
    V^r_{n'L'J'M';nLJM}=&\,-ie  (-1)^{1/2+J+L'+1} h_{pw} C^{J'M'}_{JM10}\sqrt{\frac{(2J+1)(2L+1)}{2\pi}} \sixj{J'}{J}{1}{L}{L'}{1/2}\times\\
    &\times  C^{L'0}_{L010} E_{n'L'J'M';nLJM} r_{n'L'J';nLJ}.
\end{split}
\end{equation}
Let us compare this expression with the matrix element of the operator $-d_z E^{(+)}_z(0)$, where
\begin{equation}
    \mathbf{E}^{(+)}(0)=-\dot{\spA}^{(+)}(0)=i\sum_\la \int\frac{d\spk k_0}{\sqrt{2(2\pi)^3 k_0}} \spe_{\la}(\spk) d_\la(\spk).
\end{equation}
For a plane electromagnetic wave of the form \eqref{Laguerre_Gauss} with $n_\ga=l=0$ satisfying the above assumptions, we find
\begin{equation}\label{E_+_0}
    \mathbf{E}^{(+)}(0)=\frac{ik_0}{\sqrt{2\pi}} h_{pw}\spe_{(s)}.
\end{equation}
Singling out the linear polarization $\spe_z$ in this relation and substituting it into the expression for the matrix element of the dipole moment operator \eqref{dipole_mom_matr_elem}, we arrive at
\begin{equation}
\begin{split}
    -E^{(+)}_z(0)\lan n'L'J'M'|(\spe_z \mathbf{d})|nLJM\ran =&\,-ie k_0 h_{pw} (-1)^{1/2+J+L'+1} \sqrt{\frac{(2J+1)(2L+1)}{2\pi}} \sixj{J'}{J}{1}{L}{L'}{1/2}\times\\
    &\times C^{J'M'}_{JM10} C^{L'0}_{L010} r_{n'L'J';nLJ},
\end{split}
\end{equation}
which is in agreement with \eqref{V_dip_expl} in the vicinity of the resonance $k_0=E_{n'L'J'M';nLJM}$. Furthermore, expression \eqref{E_+_0} allows one to connect $h_{pw}$ with the intensity of the coherent source of plane electromagnetic waves at the center of the Gaussian beam, i.e., at the point where it reaches its maximum. As long as
\begin{equation}
    I_m=2\mathbf{E}^{(+)}(0)[\mathbf{E}^{(+)}(0)]^*=\frac{k_0^2}{\pi}|h_{pw}|^2,
\end{equation}
we have
\begin{equation}
    |h_{pw}|=\frac{\sqrt{\pi I_m}}{k_0}.
\end{equation}
In the next section, we will use expressions \eqref{matr_elem_3}, \eqref{V_matr_elem}, \eqref{h_pjm_app}, \eqref{E1_long_wavelngth}, \eqref{Ej_long_wavelngth}, and \eqref{V_dip_expl} to describe the dynamics of an outer electron in an alkali atom.

\section{Detector of twisted radiowaves}\label{Detectors_Twisted_Radiowaves}

\subsection{Bloch equations}\label{Bloch_Eqs}

Let us find the explicit expression for the optical Bloch equations for the outer electron in an alkali atom in the resonance approximation. The stationary solution to this system of equations allows one to obtain the dielectric susceptibility of a gas of alkali atoms.

As it will be discussed below, in order to detect unambiguously the twisted radiowaves, it is necessary to remove the degeneracy with respect to the magnetic quantum number by placing the alkali atoms into the external magnetic field. Therefore, we will number the states of the outer electron in the alkali atom by two indices $(a,\s)$, where $a=(nLJ)$ and $\s=M$. The electron states differing only by the values of $\s$ are assumed to have close energies. The dynamics of the density matrix of the outer electron in the atom are approximately described by the equation \cite{Happer2010,Cohen-Tannoudji_book_2004}
\begin{equation}\label{Bloch_eqn}
    \dot{\rho}_{a\s,b\s'}=-i[h,\rho]_{a\s,b\s'} +L_{a\s,b\s'}[\rho],
\end{equation}
where the dot means the derivative with respect to time,
\begin{equation}
    h_{a\s,b\s'}=E_{a\s}\de_{ab}\de_{\s\s'} +\sum_{r=1}^{N_L} \big(V^r_{a\s,b\s'} e^{-ik_0^rt} +V^{r\dag}_{a\s,b\s'} e^{ik_0^rt} \big),
\end{equation}
and $E_{a\s}$ is the energy of the electron state $(a,\s)$. The dissipative contribution to the Bloch equation has the standard form \cite{Happer2010,Cohen-Tannoudji_book_2004,Downes2023}. The diagonal terms of this contribution are nothing but the right-hand side of the balance equation
\begin{equation}\label{L_dec_diag}
    L_{a\s,a\s}=\sum_{(b,\s')\neq(a,\s)} \rho_{b\s',b\s'} \Ga_{b\s',a\s} - \rho_{a\s,a\s}\Ga_{a\s},
\end{equation}
where $\Ga_{b\s',a\s}$ is the probability of transition from the state $(b,\s')$ to the state $(a,\s)$ per unit time due to spontaneous radiation of a photon and
\begin{equation}
    \Ga_{a\s}=\sum_{(b,\s')\neq(a,\s)} \Ga_{a\s,b\s'}.
\end{equation}
Henceforth, we assume that $\Ga_{a\s,b\s'}=0$ for $E_{a\s;b\s'}:=E_{a,\s}-E_{b\s'}<0$. In particular, $\Ga_{a\s}=0$ for the ground state. The dissipative part out of the diagonal is
\begin{equation}\label{L_nondiag}
    L_{a\s,b\s'}=-\frac12(\Ga_{a\s}+\Ga_{b\s'})\rho_{a\s,b\s'}+L^{deph}_{a\s,b\s'},
\end{equation}
where $L^{deph}_{a\s,b\s'}$ takes effectively into account the finite linewidths of coherent sources of electromagnetic waves irradiating the atom. The contributions \eqref{L_dec_diag} and the first term in \eqref{L_nondiag} account effectively for the interaction of the electron with the quantum electromagnetic field. Such an effective approach is justified since the Bloch equations describe the evolution of density matrix of the electron in the atom on time scales much larger than the formation time of spontaneous radiation. It is clear that Eq. \eqref{Bloch_eqn} respects the normalization condition
\begin{equation}\label{norm_cond}
    \sum_{a,\s} \rho_{a\s,a\s}=1,
\end{equation}
the Hermiticity, and positive-definiteness of the density matrix.

Further, we suppose that the states $a$ are numbered in an ascending order of energy beginning with $1$ and $a$ is the number of this state. Besides, we assume that the frequencies of the coherent sources of radiation $k_0^a$ are close to the transition energies $E_{a+1,\s;a\s'}=E_{a+1,\s}-E_{a\s'}$. Let $\ga_a$ be the linewidth of the radiation sources with frequency $k_0^a$. Then \cite{Downes2023,Gea-Banacloche1995}
\begin{equation}
    L^{deph}_{a\s,b\s'}=-(\ga_{a}+\ga_{a-1}+\cdots+\ga_{b-1})\rho_{a\s,b\s'},\quad a>b.
\end{equation}
For $a<b$, we have $L^{deph}_{a\s,b\s'}=(L^{deph}_{b\s',a\s})^*$.

Now we retain only those terms in the Bloch equations that are in agreement with the energy conservation law emerging at large evolution times. Bearing in mind that the term
\begin{equation}
    V^r_{a\s,b\s'} e^{-ik_0^rt}
\end{equation}
describes the absorption of a photon $\ga_r$ in the process $(b\s')+\ga_r\rightarrow (a\s)$ and the term
\begin{equation}
    V^{r\dag}_{a\s,b\s'} e^{ik_0^rt}
\end{equation}
is responsible for the emission of the photon $\ga_r$ in the process $(b\s')\rightarrow (a\s) +\ga_r$, we arrive at the approximate equation
\begin{equation}\label{Bloch_eqn_rwa_0}
\begin{split}
    \dot{\rho}_{a\s,b\s'}=&-iE_{a\s,b\s'}\rho_{a\s,b\s'} -iV_{a\s;a-1,\bar{\s}}^{a-1} e^{-ik_0^{a-1}t}\rho_{a-1,\bar{\s};b\s'} -iV_{a\s;a+1,\bar{\s}}^{a,\dag} e^{ik_0^{a}t}\rho_{a+1,\bar{\s};b\s'}+\\
    &+i\rho_{a\s;b+1,\bar{\s}} e^{-ik^{b}_0t} V^b_{b+1,\bar{\s};b\s'} +i\rho_{a\s;b-1,\bar{\s}} e^{ik^{b-1}_0t} V^{b-1,\dag}_{b-1,\bar{\s};b\s'} +L_{a\s,b\s'}[\rho],
\end{split}
\end{equation}
where the summation over $\bar{\s}$ is implied. This is the so-called resonance or rotating wave approximation. As the initial Eq. \eqref{Bloch_eqn}, Eq. \eqref{Bloch_eqn_rwa_0} preserves the normalization \eqref{norm_cond}. It is convenient to rewrite this equation in the stationary form by introducing the new variables
\begin{equation}\label{dens_matr_rwa}
    \tilde{\rho}_{a\s,b\s'}=e^{i\vf_{ab}t} \rho_{a\s,b\s'},
\end{equation}
where $\vf_{ab}:=\vf_a-\vf_b$ and
\begin{equation}
    \vf_a:=k_0^1+\cdots +k_0^{a-1},\qquad \vf_1:=0.
\end{equation}
In what follows, for brevity, we do not write the upper index at $V_{a\s;a-1,\bar{\s}}^{a-1}$ and $V_{a\s;a+1,\bar{\s}}^{a,\dag}$. Then
\begin{equation}\label{Bloch_eqn_rwa}
\begin{split}
    \dot{\tilde{\rho}}_{a\s,b\s'}= &-i\tilde{E}_{a\s;b\s'} \tilde{\rho}_{a\s,b\s'} -iV_{a\s;a-1,\bar{\s}} \tilde{\rho}_{a-1,\bar{\s};b\s'} -iV_{a\s;a+1,\bar{\s}}^{\dag} \tilde{\rho}_{a+1,\bar{\s};b\s'}+\\
    &+i\tilde{\rho}_{a\s;b+1,\bar{\s}} V_{b+1,\bar{\s};b\s'} +i\tilde{\rho}_{a\s;b-1,\bar{\s}} V^{\dag}_{b-1,\bar{\s};b\s'} +L_{a\s,b\s'}[\tilde{\rho}],
\end{split}
\end{equation}
where $\tilde{E}_{a\s;b\s'}=\tilde{E}_{a\s} -\tilde{E}_{b\s'}$ and
\begin{equation}
    \tilde{E}_{a\s}=E_{a\s}-\vf_a=E_{a\s}-k_0^1-\cdots-k_0^{a-1}= E_{1\s_1}-\De^1_{\s_2\s_1}-\cdots-\De^{a-2}_{\s_{a-1}\s_{a-2}} -\De^{a-1}_{\s\s_{a-1}}.
\end{equation}
Here we have introduced the detunings of frequencies of coherent radiation sources $\De^r_{\s'\s}:=k_0^r-E_{r+1,\s';r\s}$.

After a lapse of time, the electron in atom goes to a stationary state described by the density matrix $\tilde{\rho}_{a\s,b\s'}=\bar{\rho}_{a\s,b\s'}$, which is the stationary solution to Eqs. \eqref{Bloch_eqn_rwa}. The equations for the stationary density matrix can be cast into the form
\begin{equation}\label{eqn_for_dens_matr}
\begin{split}
    \sum_{b>a} \bar{\rho}_{b\s',b\bar{\s}}i\Ga_{b\bar{\s},a\s} +V_{a\s;a-1,\bar{\s}} \bar{\rho}_{a-1,\bar{\s};a\s} +V^{\dag}_{a\s;a+1,\bar{\s}} \bar{\rho}_{a+1,\bar{\s};a\s}+&\\
    -\bar{\rho}_{a\s;a+1,\bar{\s}} V_{a+1,\bar{\s};a\s} -\bar{\rho}_{a\s;a-1,\bar{\s}} V^{\dag}_{a-1,\bar{\s};a\s}&= \bar{\rho}_{a\s,a\s}i\Ga_{a\s},\\
    \bar{\rho}_{a,b+1}V_{b+1,b} +\bar{\rho}_{a,b-1}V^{\dag}_{b-1,b} -V_{a,a-1}\bar{\rho}_{a-1,b} -V^{\dag}_{a,a+1}\bar{\rho}_{a+1,b} &= Z_{ab}\times\bar{\rho}_{ab},\quad (a,\s)\neq (b,\s'),
\end{split}
\end{equation}
where, as above, the summation over $\bar{\s}$ is understood, the matrix multiplication with respect to indices $\s$, $\s'$ is implied, and the following notation has been introduced
\begin{equation}
    Z_{a\s,b\s'}:=\tilde{E}_{a\s;b\s'} -iX_{a\s,b\s'},\qquad X_{a\s,b\s'}:=\frac12(\Ga_{a\s}+\Ga_{b\s'}) +\sum_{r=b-1}^a \ga_r.
\end{equation}
This expression for the matrix $X_{a\s,b\s'}$ is valid for $a>b$. If $a<b$, then $X_{a\s,b\s'}=X_{b\s',a\s}$ by definition. For $a=b$ and $\s\neq\s'$, the last term in the expression for $X_{a\s,b\s'}$ is absent. Moreover, the notation has been introduced
\begin{equation}
    (Z_{ab}\times\bar{\rho}_{ab})_{\s\s'}:= Z_{a\s,b\s'} \bar{\rho}_{a\s,b\s'}.
\end{equation}
In the general case, the size of the system of equations \eqref{eqn_for_dens_matr} grows rapidly in increasing the number of electron levels.

Later on (see Sec. \ref{Dielectric_Suscept_Sec}), we will need only $\bar{\rho}_{2\s,1\s'}$ of all the elements of the density matrix. These elements of the density matrix determine the dielectric susceptibility of a gas of atoms for the probe laser that possesses the frequency $k_0^1\approx E_{2\s;1\s'}$. If the power of the probe laser is small, viz., $|V_{2\s,1\s'}|$ is sufficiently small, then one can deduce the approximate solution to \eqref{eqn_for_dens_matr} for the elements of the stationary density matrix $\bar{\rho}_{2\s,1\s'}$. In the linear order with respect to $|V_{2\s,1\s'}|$, only the elements of the density matrix $\bar{\rho}_{a\s,1\s'}$ prove to be nonzero. In that order it is sufficient to solve the equations on the second line in \eqref{eqn_for_dens_matr} for $b=1$, where all the terms containing $V_{2\s,1\s'}$ are discarded for $a\geqslant3$. The corresponding solution is provided by the recurrence formula given below.

Introduce the set of superoperators acting on the elements of the density matrix by the rule
\begin{equation}
\begin{split}
    M^{(k)}_{n_l-k,1}[\bar{\rho}_{n_l-k,1}]&=Z_{n_l-k,1}\times\bar{\rho}_{n_l-k,1}- V^{\dag}_{n_l-k,n_l-k+1} (M^{(k-1)}_{n_l-k+1,1})^{-1}[V_{n_l-k+1,n_l-k}\bar{\rho}_{n_l-k,1}],\\
    M^{(0)}_{n_l1}[\bar{\rho}_{n_l1}]&=Z_{n_l1}\times \bar{\rho}_{n_l1},
\end{split}
\end{equation}
where $k=\overline{0,n_l-2}$ and $n_l=N_L+1$ is the number of electron states disregarding the magnetic quantum number. Then the following recurrence relation holds
\begin{equation}
    \bar{\rho}_{a1}=-(M^{(n_l-a)}_{a1})^{-1}[V_{a,a-1}\bar{\rho}_{a-1,1}].
\end{equation}
Consequently,
\begin{equation}\label{rho21}
    \bar{\rho}_{21}=-(M^{(n_l-2)}_{21})^{-1}[V_{21}\bar{\rho}_{11}].
\end{equation}
In the particular case where the splitting of electron levels with respect to the magnetic quantum number is absent, i.e., the index $\s$ takes only one value, expression \eqref{rho21} coincides with the known solution to the stationary Bloch equations \cite{Downes2023,McGloin2001,Pandey2013}. Formula \eqref{rho21} allows one to express $\bar{\rho}_{2\s,1\s'}$ through the elements of the density matrix $\bar{\rho}_{1\s,1\s'}$. Substituting this expression into the second equation in \eqref{eqn_for_dens_matr} for $a=b=1$, one derive $\bar{\rho}_{1\s,1\s'}$ for $\s\neq\s'$ in terms of $\bar{\rho}_{1\s,1\s}$. The matrix elements $\bar{\rho}_{1\s,1\s'}$ for $\s\neq\s'$ turn out to be of the second order in $|V_{2\s,1\s'}|$. The system of equations \eqref{eqn_for_dens_matr} with the normalization condition \eqref{norm_cond} allows one to obtain $\bar{\rho}_{1\s,1\s}$ and, consequently, to find $\bar{\rho}_{2\s,1\s'}$ by using formula \eqref{rho21}. However, to this end one has to solve the stationary Bloch equations in the second order in $|V_{2\s,1\s'}|$ and the solution proves to be cumbersome.

\subsection{Dielectric susceptibility}\label{Dielectric_Suscept_Sec}

Let us derive the dielectric susceptibility of a gas of atoms, the density matrix of electrons in these atoms having the form \eqref{dens_matr_rwa}, where $\tilde{\rho}_{a\s,b\s'}=\bar{\rho}_{a\s,b\s'}$ is the stationary solution of Eq. \eqref{Bloch_eqn_rwa}. The average dipole moment of the atom is equal to
\begin{equation}
    \lan\mathbf{d}\ran=\sum_{(a,\s),(b,\s')} \rho_{a\s,b\s'} \lan b\s'|\mathbf{d}|a\s\ran = \sum_{(a,\s),(b,\s')} \bar{\rho}_{a\s,b\s'} e^{-i\vf_{ab}t} \lan b\s'|\mathbf{d}|a\s\ran.
\end{equation}
Then we single out the contribution of oscillations with the frequency $k_0^1$ that is close to the transition frequencies between the electron states $(1,\s)$ and $(2,\s')$. Therefore, the polarizability of a gas of atoms at this frequency is
\begin{equation}\label{polarizability}
    \mathbf{P}=n_0 \lan\mathbf{d}\ran= n_0 \big( e^{-i k_0^1t} \sum_{\s,\s'}\bar{\rho}_{2\s,1\s'} \lan 1\s'|\mathbf{d}|2\s\ran +c.c.\big),
\end{equation}
where $n_0$ is the concentration of atoms in a gas. The strength of the electric field in the electromagnetic wave can be written as
\begin{equation}\label{ext_em_field}
    \mathbf{E}= \mathbf{E}^{(+)}_0 e^{-ik_0^1t}+ \mathbf{E}^{(-)}_0 e^{ik_0^1t},
\end{equation}
where it has been assumed that the electromagnetic wave is linearly polarized along the $z$ axis. As long as the gas of atoms is isotropic to a good accuracy,
\begin{equation}\label{chi_defn}
    \mathbf{P}^{(+)}=\chi \mathbf{E}^{(+)}.
\end{equation}
Substituting the gas polarizability \eqref{polarizability} and the expression for the electric field \eqref{ext_em_field} into the  $z$ component of the relation \eqref{chi_defn}, we come to
\begin{equation}\label{dielectric_suscept}
    \chi=\frac{n_0}{E^{(+)}_{0z}}\sum_{\s,\s'}\bar{\rho}_{2\s,1\s'} \lan 1\s'|d_z|2\s\ran,
\end{equation}
where, as it has been shown in the previous section,
\begin{equation}\label{E_+_0_hpw}
    E^{(+)}_{0z}=\frac{ik_0}{\sqrt{2\pi}} h_{pw}.
\end{equation}
The explicit expression for the dipole matrix element is presented in \eqref{dipole_mom_matr_elem}. In the dipole approximation near the resonance, $k_0^1\approx E_{1,\s;2\s'}$, we have
\begin{equation}\label{V_through_d}
    V_{2\s',1\s}=-E^{(+)}_{0z}  \lan 2\s'|d_z|1\s\ran,
\end{equation}
and so
\begin{equation}
    \chi=-\frac{n_0}{|E^{(+)}_{0z}|^2} \sum_{\s,\s'}\bar{\rho}_{2\s,1\s'} V^{\dag}_{1\s',2\s}= -\frac{n_0}{V_{2\s',1\s}} \sum_{\bar{\s},\bar{\s}'}\bar{\rho}_{2\bar{\s},1\bar{\s}'} \lan 1\bar{\s}'|d_z|2\bar{\s}\ran \lan 2\s'|d_z|1\s\ran.
\end{equation}
In the last equality, we have expressed the strength of the electric field with the aid of the relation \eqref{V_through_d}. In accordance with the Bouguer law, the intensity of the probe laser passed through the cell with alkali atoms becomes
\begin{equation}\label{intenstity_det}
    I=I_0T,\qquad T:= \exp(-k^1_0L_{cuv} \im\chi),
\end{equation}
for $\im\chi\ll1$, where $T$ is the transmission coefficient, $L_{cuv}$ is the length of the cuvette along the laser ray with the frequency $k_0^1$, and $I_0$ is the initial intensity of the laser.

It has been shown in the previous section that the signal of the local oscillator (heterodyne) can be taken into account by replacing the power of the emitter in the expression for $h^{pj}_{m_\ga}$ by the effective complex power \eqref{P_eff_expl}. It is not difficult to see that the dielectric susceptibility \eqref{dielectric_suscept} depends only on the modulus of this effective power \eqref{P_eff_modul}. Indeed, as follows from \eqref{h_pjm_app_heter}, the matrix elements of the interaction Hamiltonian \eqref{V_matr_elem_rotated} entering into the Bloch equations \eqref{Bloch_eqn_rwa} depend on the phase $2\psi_{a}$ of the effective power $P_{eff}^a$ of the source with number $a$ as
\begin{equation}
    V_{a+1,\s;a\s'}= \tilde{V}_{a+1,\s;a\s'} e^{i\psi_{a}},
\end{equation}
where $\tilde{V}_{a+1,\s;a\s'}$ depends only on $|P_{eff}^a|$. Then, on making the substitution,
\begin{equation}\label{rho_prime}
    \tilde{\rho}_{a\s;b\s'} \rightarrow \tilde{\rho}_{a\s,b\s'}e^{i(f_b-f_a)},
\end{equation}
where
\begin{equation}
    f_a=-\sum_{r=1}^{a-1}\psi_r,\qquad f_1=0,
\end{equation}
we see that the solution to the Bloch equations \eqref{Bloch_eqn_rwa} after this substitution ceases to depend on the phases of the effective powers of the sources. Replacing $\bar{\rho}_{a\s,b\s'}$ in the expression for the dielectric susceptibility  \eqref{dielectric_suscept} as in \eqref{rho_prime} and bearing in mind that $E^{(+)}_{0z}\sim \exp(i\psi_1)$, we obtain that the dielectric susceptibility is independent of the phases $\psi_a$. Assuming that the change of the intensity of the laser wave \eqref{intenstity_det} passing through the vapor cell is small on switching on the information wave with power $P^a_{sig}$ and taking into account the expression for the modulus of the effective power \eqref{P_eff_modul}, we find
\begin{equation}\label{intensity_var}
    \frac{\de I}{I}\approx 2\varkappa_a\sqrt{P^a_0P^a_{sig}} \cos(\arg A_{sig}) \frac{\im\chi'}{\im\chi},
\end{equation}
where the prime means the derivative with respect to the modulus of the effective power of $a$-th source. In the absence of the local oscillator, the analogous expression reads
\begin{equation}
    \frac{\de I}{I}\approx P^a_{sig} \frac{\im\chi'}{\im\chi}.
\end{equation}
Now it is clear that the use of the reference electromagnetic wave not only increases the relative sensitivity but also allows one to measure the cosine of the relative phase between the local oscillator and the information electromagnetic wave \cite{Simons2019}. Further, we will consider the case where the information signal is given by the source of the electromagnetic waves with number $N_L$, i.e., $a=N_L$ in expression \eqref{intensity_var}.

\subsection{First scheme of the detector}\label{Level_System}

The scheme of the detector of twisted radiowaves employing the vapor of alkali atoms is analogous to the scheme of the detector of plane radiowaves. The cuvette with the vapor of alkali atoms is irradiated by the probe and coupling lasers linearly polarized along the $z$ axis and propagating along the $x$ axis in opposite directions as shown in Fig. \ref{scheme_7level_plots}. The probe laser transfers the outer electron in the alkali atom from the ground state to the low-lying excited level, whereas the coupling laser transfers the electron from this level to the Rydberg state. Furthermore, this vapor cell is irradiated by the radiowaves with different frequencies that induces transitions between the Rydberg states as shown in Fig. \ref{scheme_7level_plots}. The source of twisted radiowaves inducing nondipole transitions between the Rydberg states is among these coherent sources. The radiowaves propagates along the $z$ axis and possess circular polarizations. The vapor cell is placed into the stationary homogeneous magnetic field directed along the $z$ axis that removes the degeneracy with respect to the magnetic quantum number by the Zeeman splitting. The removal of this degeneracy is necessary for the detector to record only twisted radiowaves with modulus of the projection of the total angular momentum onto the propagation axis larger than two. If this degeneracy is kept intact, then, as is seen from the matrix element of the interaction Hamiltonian \eqref{V_matr_elem}, the plane waves irradiating the cuvette with atoms at the same frequency as the twisted radiowaves excite the nondipole transitions of any multipolarity $2^j$. In the case when the degeneracy with respect to the magnetic quantum number is eliminated, the selection rules following from the Clebsch-Gordan coefficient in \eqref{V_matr_elem} prohibit the transitions that change the magnetic quantum number by a quantity larger than one for the one-photon processes with plane electromagnetic waves.

\begin{table}[t]
\begin{center}
  \begin{tabular}{ | l | c | c | c | c | c | }
    \hline
      & $62\prescript{2}{}{S}_{1/2}$ & $62\prescript{2}{}{P}_{3/2}$ & $61\prescript{2}{}{D}_{5/2}$ & $59\prescript{2}{}{F}_{7/2}$ & $59\prescript{2}{}{H}_{11/2}$\\ \hline
    $E$, GHz & -979.620 & -963.254 & -960.200 & -946.151 & -945.161 \\ \hline
    $\De E$, MHz & 140 & 93.4 & 84.0 & 80.0  & 76.4  \\
    \hline
  \end{tabular}
\end{center}
\caption{{\footnotesize The energies of the unperturbed Rydberg states and the energy difference between the neighboring Zeeman states in the magnetic field with the strength $H_z=50$ G. }}\label{Tab_energy_split}
\end{table}

As discussed in Sec. \ref{Matr_Elem_H_Int_CM}, in order to neglect the form of the wave function of the center of mass and, in particular, to neglect the transfer of the angular momentum from the electromagnetic field to the center of mass \cite{Mukherjee2018,Peshkov2023}, it is necessary that the twisted radiowaves obey the estimate \eqref{estim_twisted}. Since the size of the vapor cell is of order of several centimeters, the wavelength of the nondipole transition should be larger than $10$ cm, i.e., the frequency of this transition should be smaller than $3$ GHz. Both for cesium and rubidium such nondipole transitions with $\De L=2$ exist either for the transitions with the principal quantum numbers near $100$ or larger or for the transitions $L\rightarrow L+2$ with $L\geqslant3$ \cite{Elgee2023,Allinson2024,Sibalic2017ARC}. Below we will study the latter case for cesium. The length of the cuvette along the probe laser ray, $L_{cuv}$, is taken to be $2$ cm.

Let us consider the following level scheme
\begin{equation}\label{level_scheme}
    6\prescript{2}{}{S}_{1/2}(F=4)\rightarrow 6\prescript{2}{}{P}_{1/2}(F=4)\rightarrow 62\prescript{2}{}{S}_{1/2}\rightarrow 62\prescript{2}{}{P}_{3/2}\rightarrow 61\prescript{2}{}{D}_{5/2} \rightarrow 59\prescript{2}{}{F}_{7/2}\rightarrow 59\prescript{2}{}{H}_{11/2}.
\end{equation}
All the transitions in this chain, apart from the last one, are $E1$-transitions. The last transition is $E2$. We suppose that the atom is located in the external magnetic field directed along the $z$ axis with the field strength of order $50$ G or less. Then the external magnetic field perturbs weakly the hyperfine structure of the levels $6\prescript{2}{}{S}_{1/2}(F=4)$ and $6\prescript{2}{}{P}_{1/2}(F=4)$ and the splitting of energy of the unperturbed level is described with good accuracy by \cite{Frish1963,LandLifshQM.11,Steck2024Cs,Steck_book_2025}
\begin{equation}
    \De E =\mu_B g_F m_F H_z,\qquad m_F=\overline{-F,F},
\end{equation}
where
\begin{equation}
    g_F=g_J \frac{F(F+1)+J(J+1)-I(I+1)}{2 F(F+1)},\qquad g_J=1+\frac{J(J+1)+S(S+1)-L(L+1)}{2J(J+1)},
\end{equation}
and $S=1/2$, $I=7/2$ for cesium. If $H_z=50$ G, then the total magnetic splitting of the level $6\prescript{2}{}{S}_{1/2}(F=4)$ amounts to $157$ MHz and the complete magnetic splitting of the level $6\prescript{2}{}{P}_{1/2}(F=4)$ is equal to $52.5$ MHz. The energy of the unperturbed state $6\prescript{2}{}{S}_{1/2}(F=4)$ is $-941542.216$ GHz and the energy of the unperturbed state  $6\prescript{2}{}{P}_{1/2}(F=4)$ equals $-606426.167$ GHz. The energies of the first and second transitions in the scheme \eqref{level_scheme} correspond approximately to the lasers with wavelengths $894.6$ nm and $495.2$ nm, respectively. The consideration of the hyperfine structure split by the external magnetic field is rather cumbersome for the system of levels \eqref{level_scheme}. Therefore, we will simulate the system of levels for the terms $6\prescript{2}{}{S}_{1/2}(F=4)$ and $6\prescript{2}{}{P}_{1/2}(F=4)$ by two active levels for every term as in the case of accounting for the fine structure only. The energy difference between the two levels of every term is taken to be equal to the total magnetic splitting of this term given above.

\begin{table}[t]
\begin{center}
  \begin{tabular}{ | l | c | c | c | c | c | c | }
    \hline
      & $(2,1)\rightarrow(1,1)$ & $(3,1)\rightarrow(2,1)$ & $(4,1)\rightarrow(3,1)$ & $(5,1)\rightarrow(4,1)$ & $(6,1)\rightarrow(5,1)$ & $(7,1)\rightarrow(6,1)$\\ \hline
    $\De E$, GHz & 335116.101 & 605446.503 & 16.297 & 2.984 & 13.979 & 0.850\\ \hline\hline
      & $(2,2)\rightarrow(1,2)$ & $(3,1)\rightarrow(2,2)$ & $(4,2)\rightarrow(3,2)$ & $(5,2)\rightarrow(4,2)$ & $(6,2)\rightarrow(5,2)$ & $(7,2)\rightarrow(6,2)$\\ \hline
    $\De E$, GHz & 335115.996 & 605446.590 & 16.250 & 2.974  & 13.975  & 0.847\\
    \hline
  \end{tabular}
\end{center}
\caption{{\footnotesize The transition energies between the states $(a,\s)$ of the outer electron. The index $\s$ runs the values $\{1,2\}$, the value $\s=1$ corresponding to the state with the minimum energy at a given $a$. The scheme of levels is presented in Fig. \ref{scheme_7level_plots}}. }\label{Tab_energy_trans}
\end{table}

The magnetic moment of the electron in the Rydberg state interacts weakly with the magnetic moment of the nucleus. Consequently, the hyperfine structure can be neglected for these states (see, however, \cite{Ripka2022}). The splitting of energy of the unperturbed Rydberg level in the magnetic field is described by the formula
\begin{equation}
    \De E =\mu_B g_J m_J H_z,\qquad m_J=\overline{-J,J}.
\end{equation}
The differences between the neighboring Zeeman levels in the magnetic field are presented in Table \ref{Tab_energy_split}. We suppose that the emitters of radiowaves inducing the transitions between the Rydberg states create the electromagnetic fields possessing the circular polarization and propagating along the $z$ axis. They cause $E1$-transitions with $\De m_J=-1$ and $E2$-transition with $\De m_J=-2$. As a result, only the two lower levels of the every term $nLJ$ with minimal $m$ and energies become populated (see Fig. \ref{scheme_7level_plots}). Therefore one can take $\s$ in the Bloch equations \eqref{Bloch_eqn_rwa} to run only two values: $\sigma=1$ denotes $m_J = -J$ and $\sigma=2$ denotes $m_J = -J+1$.

The probability of radiation per unit time for the $2^j$-pole transition $nLJ\rightarrow n'L'J'$ summed and averaged with respect to the magnetic quantum numbers of the initial and the final states and summed over the magnetic quantum numbers of the emitted photon equals
\begin{equation}
    \Ga_{nLJ;n'L'J'}=\frac{1}{2J+1}\sum_{M',M,m} \Ga_{nLJM;n'L'J'M'}(m).
\end{equation}
The quantities $\Ga_{nLJ;n'L'J'}$ can be easily found by using the software package (ARC). It follows from the Wigner-Eckart theorem that the dependence of $\Ga_{nLJM;n'L'J'M'}(m)$ on $M'$, $M$, and $m$ is determined by the factor $(C^{J'M'}_{JMjm})^2$. Consequently, the probability of radiation per unit time for the $2^j$-pole transition $nLJM\rightarrow n'L'J'M'$ summed over the magnetic quantum numbers of the escaped photon (for the dipole transition this summation is equivalent to summation over the photon polarizations) has the form
\begin{equation}
    \Ga_{nLJM;n'L'J'M'}=\frac{2J+1}{2J'+1}\sum_{m=-j}^{j}(C^{J'M'}_{JMjm})^2 \Ga_{nLJ;n'L'J'}=\frac{2J+1}{2J'+1}(C^{J'M'}_{JMj,M'-M})^2 \Ga_{nLJ;n'L'J'}.
\end{equation}
It is not difficult to see that
\begin{equation}
    \sum_{M'}\Ga_{nLJM;n'L'J'M'}= \Ga_{nLJ;n'L'J'}.
\end{equation}
Therefore, the decay rate of the state $(a,\s)$ entering into the dissipative part \eqref{L_nondiag} of the Bloch equations does not depend on $\s$.

\begin{figure}[tp]
\centering
\includegraphics*[width=0.45\linewidth]{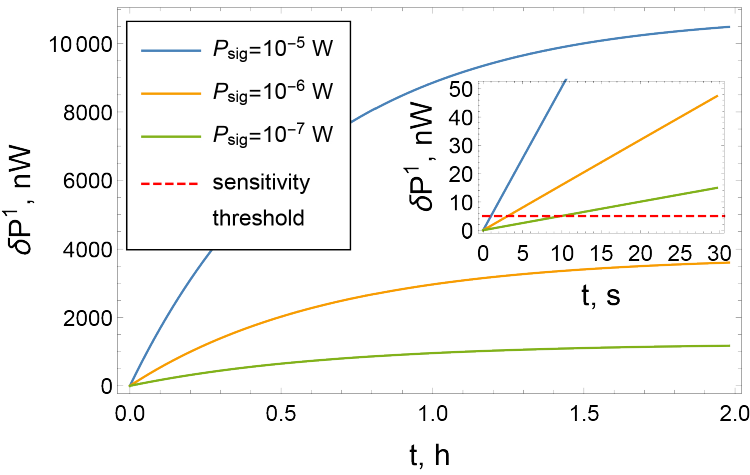}\;
\includegraphics*[width=0.45\linewidth]{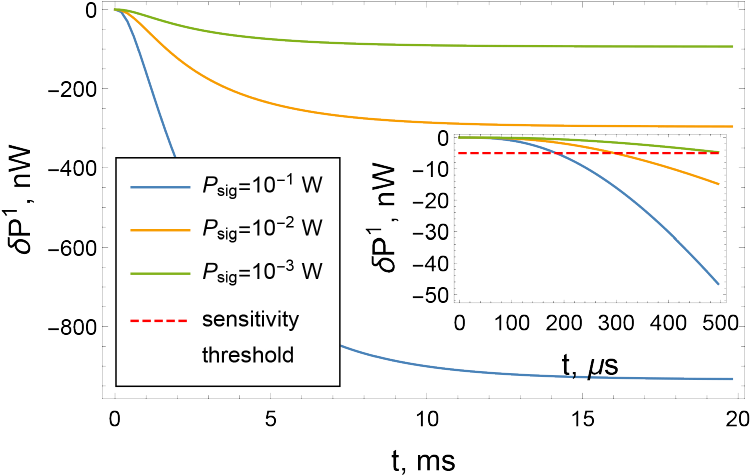}
\caption{{\footnotesize The dynamics of the power change of the probe laser measured by the photodetector for the Rydberg-atom based receiver described in Fig. \ref{scheme_7level_plots}. We put $\vk_6\cos(\arg A_{sig})=1$. The sensitivity threshold is taken to be $5$ nW. On the left panel: The powers of coherent sources are chosen as in \eqref{powers_sources}. On the right panel: The powers of coherent sources are presented in \eqref{powers_sources_perf}.}}
\label{power_change_7level_plots}
\end{figure}

Thus we have all the necessary ingredients to solve the Bloch equations \eqref{Bloch_eqn_rwa}. The results of the numerical solution of this system of equations are given in Fig. \ref{power_change_7level_plots}. The parameters for the numerical simulations are taken as follows. The temperature of the atomic gas determining the concentration of cesium atom in the cell and influencing the transition rates of the electron in the atom is taken to be $300$ K. The detunings of coherent sources are put to zero. A small change of the deturnings shows that the deviation from the resonances leads to decreasing of the detector sensitivity. The linewidths of the coherent sources are $10$ kHz for the probe and the coupling lasers and $1$ kHz for the sources of radiowaves. The standard deviation of the perpendicular component of the photon momentum, $\s_\perp$, is set to $119.3$ GHz for the probe and coupling lasers that corresponds approximately to $400$ $\mu$m for the beam waist in the coordinate space. As for the sources of plane radiowaves and twisted radiowaves, we take $\s_\perp=100$ kHz and $\s_\perp=100$ MHz, respectively. The transition probabilities per unit time and the radial integrals are calculated with the help of the package \cite{Sibalic2017ARC}. The powers of sources \eqref{power_expl} entering into \eqref{h_pjm_app} are chosen to maximize the absolute sensitivity of detector with account for the presence of the heterodyne, viz., the quantity (see \eqref{intenstity_det}, \eqref{intensity_var})
\begin{equation}
    \eta:=\sqrt{P_0^6}|T'| P^1
\end{equation}
is maximized, where the prime denotes the derivative with respect to the power $P_0^6$ of the local oscillator generating the twisted radiowaves. The numerical optimization gives rise to the following values for the powers of sources
\begin{equation}\label{powers_sources}
    P^a=\{6.22\times 10^{-2}, 6.37\times 10^{-1}, 3.21\times 10^{-1}, 5.42\times 10^{-1}, 2.53\times 10^{-2}, 3.61\times 10^{-5}\}\;\text{W}.
\end{equation}
These values of powers lead to
\begin{equation}
    \eta=2.61\times10^{-3}\;\text{W}^{1/2}.
\end{equation}
For example, for the power of twisted information electromagnetic wave $P_{sig}^6=1$ $\mu$W, the power change of the probe laser on the photodetector is of order $\de P^1=5$ $\mu$W, where it is assumed that $\varkappa_6\cos(\arg A_{sig})\approx1$. In decreasing the power of the information signal $P_{sig}^6$, the power change of the probe signal $\de P^1$ on the photodetector decreases as $(P_{sig}^6)^{1/2}$. Keeping in mind that the threshold sensitivity of photodetectors is of order $5$ nW, the detector of twisted radiowaves we consider allows one to record the twisted signal of a very small power. In order to find the precise estimate for the lower bound of the signal power that can be registered by this detector, it is necessary to take into account the influence of the Doppler broadening, the cell geometry, and other parasitic effects. The results of the papers \cite{Liu2022,Gea-Banacloche1995,Wen2024} show that the approach we use is a good zeroth order approximation to catch the main physical effects observed experimentally. Moreover, the Doppler broadening can be reduced by adding the auxiliary level and the corresponding laser source \cite{Ryabtsev2011,Ripka2022,Sandhya1997,Fulton1995,Gea-Banacloche1995}. We leave the investigation of these effects for future research.

Despite a high sensitivity, one of the drawbacks of this detector of twisted radiowaves for the parameters given above is the large time to enter the steady state. In Fig. \ref{power_change_7level_plots}, we present the dynamics of the power change of the photodetector when the system being in the steady state is perturbed by the information wave with the power $P^6_{sig}$ and tends to the new steady state. For example, for $P^6_{sig}=1$ $\mu$W, the power change of the probe signal $\de P^1$ on the photodetector reaches the value $45$ nW in $30$ s. Taking the other values of the powers of sources \eqref{powers_sources}, one can increase the operating speed of the detector at the cost of decreasing its sensitivity. For example, for the powers of sources
\begin{equation}\label{powers_sources_perf}
    P^a=\{5.81\times 10^{-3}, 7.82\times 10^{-2}, 1.97\times 10^{-2}, 7.47, 1.42\times 10^{-3}, 26.1\}\;\text{W},
\end{equation}
and $P^6_{sig}=100$ mW, $\de P^1$ reaches $48$ nW in $0.5$ ms that corresponds to the frequency of the transmitted information signal $2$ kHz.

\subsection{Second scheme of the detector}\label{Second_Scheme_Det}

\begin{figure}[tp]
\centering
\includegraphics*[width=0.8\linewidth]{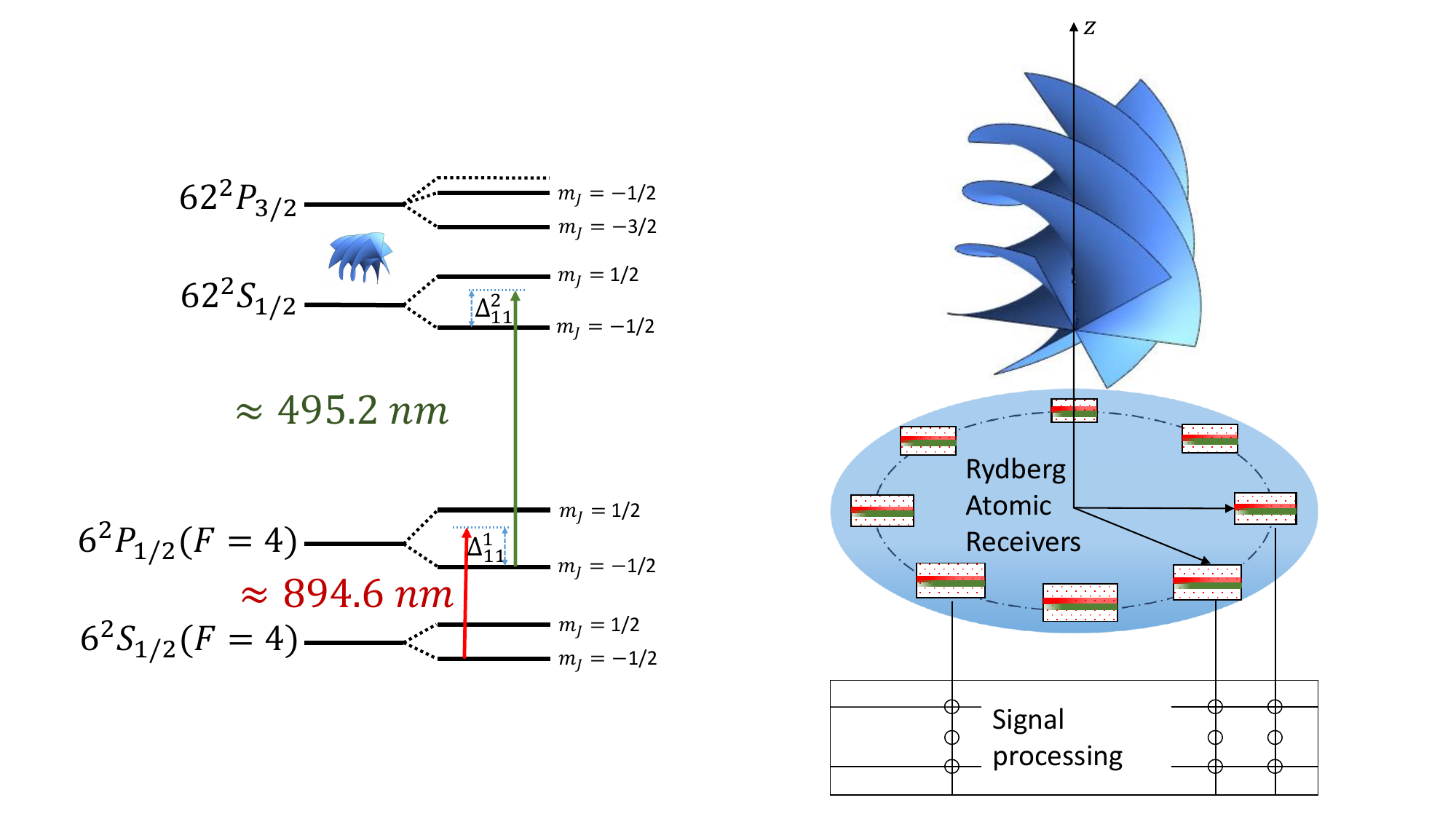}
\caption{{\footnotesize The scheme of the Rydberg-atom based detector of twisted radiowaves described in Sec. \ref{Second_Scheme_Det}. The signal of twisted radiowaves is heterodyned.}}
\label{scheme_4level_plots}
\end{figure}

As we have seen in Sec. \ref{Dielectric_Suscept_Sec}, the use of the heterodyne allows one to determine the cosine of the phase between the local oscillator and the information electromagnetic wave. This property enables one to record the twisted radiowaves receiving the radio signal by an array of antennas based on Rydberg atoms. Every antenna from the array records the plane radiowaves and can be standardly realized by means of the four level scheme with heterodyne \cite{Gong2025,Richardson2025,Cui2025,Kim2025,Gong2024}. The magnetic field for removal of degeneracy with respect to the magnetic quantum number is not needed in this case. The antennas are conveniently placed on a circle or on its arc (see Fig. \ref{scheme_4level_plots}). Then, knowing the phase difference, $\arg A_{sig}$, between the reference and information waves, one can restore the information signal multiplexed by the projection of the angular momentum. Several procedures of such a multiplexing with beam forming are considered in detail in \cite{annalsOAMmult} (see also \cite{Zheng2015,Zheng2022,Chen2022,YaSaLee22}). In order to find $\arg A_{sig}$ and not only $\cos(\arg A_{sig})$, one can abruptly change the phase of the local oscillator by $\pi/2$ in receiving the information wave. This allows one to measure $\cos(\arg A_{sig})$ and $\sin(\arg A_{sig})$ by measuring the intensity of the probe laser on the photodetector (see formulas \eqref{intensity_var}).

\begin{figure}[tp]
\centering
\includegraphics*[width=0.45\linewidth]{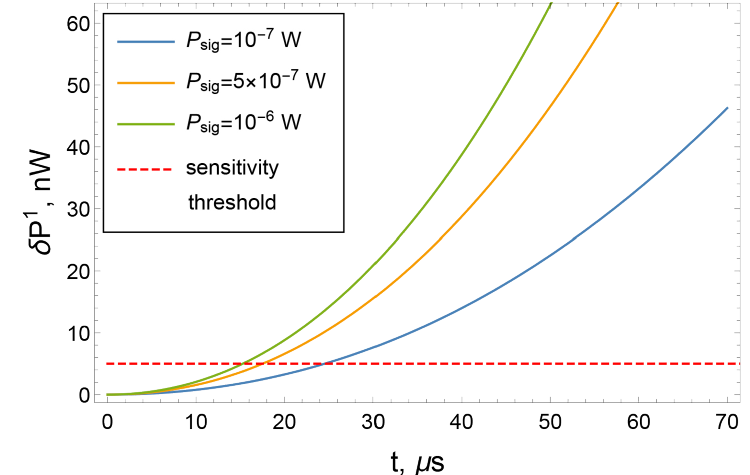}
\caption{{\footnotesize The dynamics of the power change of the probe laser measured by the photodetector for the Rydberg-atom based receiver described in Fig. \ref{scheme_4level_plots}. We put $\vk_3\cos(\arg A_{sig})=1$. The sensitivity threshold is taken to be $5$ nW. The powers of coherent sources are given in \eqref{powers_sources_4level}.}}
\label{power_change_4level_plots}
\end{figure}

To compare the characteristics of the detector of twisted radiowaves described in this section with the detector studied in the previous section, we consider the level scheme \eqref{level_scheme}, where only the first four levels are retained. In that case, there are only three coherent sources of radiation: the probe laser, the coupling laser, and the source of radiowaves with frequency $16.297$ GHz. A similar level scheme was considered in \cite{Wen2024}. The source of radiowaves contains the contributions of the local oscillator and the information electromagnetic wave. The polarizations of the coherent sources of radiation, the detunings, the linewidths, and the dispersions of the perpendicular component of the photon momentum are the same as in the previous section. As in the previous section, we take into account only the fine structure of the states. The results of numerical simulations are presented in Fig. \ref{power_change_4level_plots}. The powers of sources are adjusted so as to increase the absolute sensitivity of detector with account for heterodyning, i.e., the quantity
\begin{equation}\label{eta_2}
    \eta:=\sqrt{P_0^3}|T'| P^1
\end{equation}
is maximized, where the prime means the derivative with respect to the power of reference wave $P_0^3$. For the powers of sources
\begin{equation}\label{powers_sources_4level}
    P^a=\{6.80\times 10^{-2}, 6.60\times 10^{-1}, 8.00\times10^{-7}\}\;\text{W},
\end{equation}
the quantity \eqref{eta_2} equals
\begin{equation}
    \eta=3.38\times 10^{-2}\;\text{W}^{1/2}.
\end{equation}
For example, for the power of information electromagnetic wave $P_{sig}^3=100$ nW, the power change of the probe laser registered by the photodetector is of order $\de P^1=20$ $\mu$W, where it is assumed that $\varkappa_3\cos(\arg A_{sig})\approx1$. The evolution of the power of probe laser recorded by the photodetector is given in Fig. \ref{power_change_4level_plots}. It is seen that $\de P^1$ becomes of order $46$ nW in $70$ $\mu$s. As a result, in this regime the information signal can be transmitted with the frequency $7.14$ kHz, where it has been taken into account that the phase of the local oscillator has to be changed by $\pi/2$ during the information transfer that leads to a twice decrease of the frequency of the information signal. An increase of the power of information electromagnetic wave leads to an increase of the information transfer rate due to the fact that the time needed for the system to go to the regime where $\de P^1$ is of order $50$ nW or larger becomes smaller. For example, for the power $P_{sig}^3=500$ nW, $\de P^1$ reaches $47$ nW in $50$ $\mu$s that results in the frequency of  information signal equal to $10$ kHz.

\section{Conclusion}

Let us sum up the results. We have developed a general theory of dynamics of the outer electron in alkali atoms irradiated by structured electromagnetic waves, in particular, by twisted photons. This theory has allowed us to describe the properties of the two possible schemes of the detectors of twisted radiowaves based on Rydberg atoms.

The first variant of the detectors employs the nondipole transitions of the outer electron between Rydberg states. It is described in Sec. \ref{Level_System}. We have shown that for such a scheme the wavelength of twisted radiowaves should be much larger than the size of the cuvette with alkali atoms. Since the size of the vapor cell is of order of several centimeters, the transition between electron states with large angular momentum $L$ have been considered that provide the transition energies smaller than $3$ GHz. We have investigated the detector of twisted radiowaves with the projection of the total angular momentum $|m_\ga|=2$ based on $E2$-transition. We have considered the seven level ladder scheme \eqref{level_scheme} with heterodyning (see Fig. \ref{scheme_7level_plots}). Notice that a miniaturization of the vapor cell as it was done in \cite{Giat2025} would allow for increase of the maximum frequency of detected twisted radiowaves resulting in simplification of the detector level scheme. The detector has a standard design for Rydberg-atom based receivers \cite{Sedlacek2012,Sedlacek2013,Anderson2014,Holloway2017,Simons2019,Song2019,Xue2021,Chen2025,Gong2024,Nowosielski2024,Hao2024,Ryabtsev2024,Yuan2023,Liu2022,Anderson2021AMFM,Robinson2021,Anderson2021,Stelmashenko2020,Jau2020,Gordon2019}. It consists of the cuvette with the vapor of cesium atoms irradiated by the two lasers in head-on configuration: the probe laser, whose signal is recorded by the balanced photodetector, and the coupling laser transferring the outer electron in the cesium atom to a Rydberg state. There are also three coherent sources of plane radiowaves with circular polarization $s_\ga=-1$ that transfer the outer electron to the required Rydberg state. The fourth radiowave emitter produces twisted radiowaves with $m_\ga=-2$. They are a coherent sum of the reference twisted radiowave (the local oscillator) and the information twisted radiowave. Note in passing that the formalism developed in the present paper allows one to generalize easily the scheme of the detector to register the twisted radiowaves with almost arbitrary $m_\ga$. In order to detect unambiguously the twisted radiowaves, the vapor cell is placed in the stationary homogeneous magnetic field that splits the levels with different magnetic quantum numbers. We have shown that in accordance with our theoretical model such a detector feels the information twisted radiowaves with powers down to nW scale. The price for such a sensitivity is a large response time reaching tens of seconds. Nevertheless, the parameters of the detector can be adjusted so that the response time is reduced down to $0.5$ ms but the sensitivity degrades. A further optimization seems to reduce the response time but we leave this issue for a future research.

The second scheme of the detector uses the array of Rydberg-atom based antennas and resembles the corresponding MIMO systems investigated recently \cite{Gong2025,Richardson2025,Cui2025,Kim2025,Gong2024}. This type of the detector is described in Sec. \ref{Second_Scheme_Det}. The Rydberg-atom based antennas in the array respond to plane radiowaves and are realized as the four level ladder systems with only $E1$-transitions with the common heterodyne (see Fig. \ref{scheme_4level_plots}). Switching the phase of the local oscillator by $\pi/2$ during the reception of the information signal allows one to restore the relative phases of the radiowaves registered by the Rydberg-atom based antennas. Then the developed procedures can be employed to demultiplex the information signal with respect to orbital angular momentum \cite{annalsOAMmult}. For the model we consider in Sec. \ref{Second_Scheme_Det} and the parameters adopted there, such a detector of twisted radiowaves turns out to be more high-speed, more sensitive, and more flexible than the detector described in Sec. \ref{Level_System}. In particular, the detector of the second type is able to discriminate the twisted radiowaves with rather large projections of the orbital angular momentum. The obvious drawback of such a detector is that it is bulky. Moreover, in spite of the fact that this type of the detector possesses a lesser response time, this time is only of order of several $\mu$s. This time can be further diminished with the aid of the Purcell effect by placing the vapor cell in a specially designed cavity. The Purcell effect increases the decay rates and so it diminishes the time needed for the electron to go to the steady state. Besides, the cavity can be used to increases the sensitivity of the Rydberg-atom based receivers \cite{Meyer2021,Amarloo2025}. The investigation of the influence of the Purcell effect on the properties of the detectors based on Rydberg atoms will be given elsewhere.

Thus we see that the Rydberg-atom based receivers of twisted radiowaves are very sensitive and can record the signal with power down to several nW. However, the low performance of such receivers is their bottleneck. We are going to study this issue in our future research.

\paragraph{Acknowledgments.}
We are grateful to Prof. V.A. Kagadey  for fruitful discussions of Rydberg-atom based detectors.

\appendix
\section{Multipole transitions of electron in an atom}\label{App_Mult_Trans}

To ensure consistency in notation, we give in this appendix the basic formulas for the matrix elements of the Hamiltonian of interaction of an electron in an alkali atom and an external electromagnetic field. Except for the arrangement of indices in the spherical vectors, we will use the conventions adopted in \cite{Varshalovich} and \cite{AhiezSit}.

The wave function of a system with total angular momentum $(j,m)$ consisting of two non-interacting subsystems $(1)$ and $(2)$ with angular momenta $j_1$ and $j_2$ is given by
\begin{equation}\label{ang_mom_addition}
\psi^{jm}_{j_1j_2}(x_1,x_2)=\sum_{m_1,m_2} C^{jm}_{j_1m_1j_2m_2}\psi^{(1)}_{j_1m_1}(x_1) \psi^{(2)}_{j_2m_2}(x_2),
\end{equation}
where $C^{jm}_{j_1m_1j_2m_2}$ are the Clebsch-Gordan coefficients, the sum is over all admissible projections of angular momentum, and
\begin{equation}\label{eigen_states_JJz}
\mathbf{J}_{1,2}^2\psi^{(1,2)}_{jm}=j(j+1) \psi^{(1,2)}_{jm},\qquad (J_{1,2})_z\psi^{(1,2)}_{jm}=m \psi^{(1,2)}_{jm}.
\end{equation}
The functions \eqref{ang_mom_addition} satisfy the system of equations
\begin{equation}
\begin{aligned}
\mathbf{J}^2\psi^{jm}_{j_1j_2}&=j(j+1)\psi^{jm}_{j_1j_2},&\qquad J_z\psi^{jm}_{j_1j_2}&=m\psi^{jm}_{j_1j_2},\\ \mathbf{J}_1^2\psi^{jm}_{j_1j_2}&=j_1(j_1+1)\psi^{jm}_{j_1j_2},&\qquad \mathbf{J}_2^2\psi^{jm}_{j_1j_2}&=j_2(j_2+1)\psi^{jm}_{j_1j_2}.
\end{aligned}
\end{equation}
The Clebsch-Gordan coefficients obey the orthogonality and completeness relations
\begin{equation}
\sum_{j=|j_1-j_2|}^{j_1+j_2} C^{jm}_{j_1m_1j_2m_2} C^{jm}_{j_1m'_1j_2m'_2}=\de_{m_1m_1'}\de_{m_2m_2'},\qquad \sum_{m_1,m_2} C^{jm}_{j_1m_1j_2m_2} C^{j'm'}_{j_1m_1j_2m_2}=\de_{jj'}\de_{mm'}.
\end{equation}
The Clebsch-Gordan coefficients are taken to be real. When the coordinate system is rotated, the wave functions with angular momentum $(j,m)$ are transformed as
\begin{equation}\label{rotation_Wigner}
\psi_{jm}(\spx')=\sum_{m'}\psi_{jm'}(\spx) D^j_{m'm}(\al,\be,\ga),
\end{equation}
where $\al$, $\be$ and $\ga$ are the Euler angles defining the rotation of the coordinate system and the Wigner $D$-functions have the form
\begin{equation}\label{Wigner_matr}
D^j_{mm'}(\al,\be,\ga)=e^{-im\al} d^j_{mm'}(\be) e^{-im'\ga},
\end{equation}
where $d^j_{mm'}(\be)$ are the small Wigner functions.

According to the general addition formula \eqref{ang_mom_addition}, the spherical tensors with angular momentum $(j,m)$, orbital momentum $l$, and spin $s$ are written as
\begin{equation}
Y^{js}_{lm}=\sum_{m_l,m_s} C^{jm}_{lm_lsm_s}Y_{lm_l}(\theta,\vf)\chi^s_{m_s},
\end{equation}
where $Y_{lm_l}$ are the scalar spherical functions, and $\chi^s_{m_s}$ are the basis spin functions. For spin $s=1$, we obtain the spherical vectors
\begin{equation}
\mathbf{Y}^j_{lm}(\theta,\vf)=\sum_{m_l=-l}^l\sum_{m_s=-1}^1 C^{jm}_{lm_l1m_s} Y_{lm_l}(\theta,\vf) \spe_{m_s},
\end{equation}
where
\begin{equation}\label{e_pm}
\spe_0=\spe_3,\qquad\spe_\pm=\mp\frac{1}{\sqrt{2}}(\spe_1\pm i\spe_2).
\end{equation}
The spherical vectors are used to construct a complete set of mode functions of photons with definite values of the total angular momentum $j$ and its projection $m$ (the multipoles)
\begin{equation}\label{multipoles}
\bs\psi^p_{jm}(k_0,\spx)=
\left\{
\begin{array}{ll}
j_j(k_0r)\mathbf{Y}^j_{jm}(\theta,\vf), & \hbox{$p=0$;} \\
-\sqrt{\frac{j}{2j+1}} j_{j+1}(k_0r)\mathbf{Y}^j_{j+1,m}(\theta,\vf) +\sqrt{\frac{j+1}{2j+1}}
j_{j-1}(k_0r)\mathbf{Y}^j_{j-1,m}(\theta,\vf), & \hbox{$p=1$,}
\end{array}
\right.
\end{equation}
where $j_j(k_0r)$ are the spherical Bessel functions, $p=0$ are the magnetic multipoles $Mj$, and $p=1$ are the electric multipoles $Ej$. Under the action of spatial reflections, the multipoles are transformed as
\begin{equation}
P \bs\psi^p_{jm}=(-1)^{j+1+p}\bs\psi_{jm}.
\end{equation}
The normalization condition has the form
\begin{equation}
\int d\spx \bs\psi^{p*}_{jm}(\spx) \bs\psi^{p'}_{j'm'}(\spx)=\frac{\pi}{2k_0^2}\de(k_0-k_0')\de_{jj'}\de_{mm'}\de_{pp'}.
\end{equation}
A plane wave can be expanded in term of multipoles
\begin{equation}\label{mult_expans}
\spe_{(\la)}(\spk)e^{i\spk\spx}=\sum_{j=1}^\infty \sum_{m=-j}^j\sum_{p=0,1}(i\la)^pi^j\sqrt{2\pi(2j+1)}
D^j_{m\la}(\vf_k,\theta_k,0)\bs\psi^p_{jm}(\spx),
\end{equation}
where the expression for the photon polarization vector is given in \eqref{photon_polarization}

In the long-wavelength limit, one can simplify the calculation of the matrix elements of the interaction Hamiltonian responsible for $Ej$-transitions by using the Siegert theorem \cite{AhiezSit}. In the long-wavelength limit, $k_0r\ll1$, where $r\sim r_B$ is the typical size of the region where the wave function is nonzero, the following approximate equalities hold
\begin{equation}\label{Ej_mult_IR}
\bs\psi^1_{jm}\approx \sqrt{\frac{j+1}{2j+1}} j_{j-1}(k_0 r)\mathbf{Y}^j_{j-1,m}(\theta,\vf)\approx \sqrt{\frac{j+1}{2j+1}}
\frac{(k_0 r)^{j-1}}{(2j-1)!!}\mathbf{Y}^j_{j-1,m}(\theta,\vf).
\end{equation}
In the last equality we have used the relation
\begin{equation}\label{Bessel_expns}
j_j(x)\approx x^j/(2j+1)!!,
\end{equation}
that is valid for small $x$. From the gradient formula (5.8.9) of \cite{Varshalovich},
\begin{equation}\label{grad_formula}
\nabla\big(f(r)Y_{lm}(\theta,\vf)\big)=-\sqrt{\frac{l+1}{2l+1}} \big(f'-\frac{l}{r}f\big) \mathbf{Y}^l_{l+1,m} +\sqrt{\frac{l}{2l+1}} \big(f'+\frac{l+1}{r}f\big) \mathbf{Y}^l_{l-1,m},
\end{equation}
it follows that
\begin{equation}
\nabla(r^j Y_{jm}(\theta,\vf))=\sqrt{j(2j+1)} r^{j-1}\mathbf{Y}^j_{j-1,m}(\theta,\vf).
\end{equation}
As a result, for the multipole \eqref{Ej_mult_IR} we approximately obtain
\begin{equation}
\bs\psi^1_{jm}\approx \sqrt{\frac{j+1}{j}} \frac{k_0^{j-1}}{(2j+1)!!} \nabla\big(r^jY_{jm}(\theta,\vf) \big) \approx \sqrt{\frac{j+1}{j}} \frac{1}{k_0}\nabla\big(j_j(k_0r)Y_{jm}(\theta,\vf) \big),
\end{equation}
i.e., $Ej$-multipole is a gradient in the long-wavelength limit.

The matrix element of the interaction Hamiltonian of an electron with an external classical electromagnetic field has the form
\begin{equation}
-\int d\spx  \spA(t,\spx) \lan f|\hat{\mathbf{J}}(\spx)| i\ran,
\end{equation}
where $|f\ran$ and $|i\ran$ are the initial and final states of the electron, which we assume to be the eigenstates of its Hamiltonian $\hat{H}_e$, and $\hat{\mathbf{J}}(\spx)$ is the current density operator satisfying the continuity equation
\begin{equation}\label{charge_cons_law}
\partial_i\hat{J}_i(\spx)=-i[\hat{H}_e,\hat{\rho}],
\end{equation}
where $\hat{\rho}$ is the charge density operator. It is clear from the multipole expansion \eqref{mult_expans} that the matrix element of the interaction Hamiltonian describing $Ej$-transition is proportional to
\begin{equation}
M^E_{jm}:=-\int d\spx \bs\psi^1_{jm}(\spx)\lan f|\hat{\mathbf{J}}(\spx)|i\ran \approx \sqrt{\frac{j+1}{j}}\frac{1}{k_0} \int d\spx  j_j(k_0r)Y_{jm}(\theta,\vf) \lan f| \partial_i\hat{J}_i(\spx)|i\ran.
\end{equation}
Integrating by parts and using \eqref{charge_cons_law}, we arrive at
\begin{equation}
M^E_{jm}\approx -i\sqrt{\frac{j+1}{j}} \frac{E_{fi}}{k_0} \int d\spx j_j(k_0r)Y_{jm}(\theta,\vf)\lan f|\hat{\rho}(\spx)|i\ran \equiv -i\sqrt{\frac{j+1}{j}} \frac{E_{fi}}{k_0} \lan f|M^C_{jm}|i\ran,
\end{equation}
where $E_{fi}=E_f-E_i$. For a single active electron,
\begin{equation}
\hat{\rho}(\spx)=e\de(\spx-\spx_e),
\end{equation}
we obtain
\begin{equation}\label{MEjm_IR}
M^E_{jm}\approx -ie\sqrt{\frac{j+1}{j}} \frac{E_{fi}}{k_0} \lan f|j_j(k_0r_e)Y_{jm}(\theta_e,\vf_e)|i\ran.
\end{equation}
For the electron states constructed according to the $LS$-coupling scheme, i.e., as in \eqref{ang_mom_addition}, the integral over angular variables in the appearing matrix element can be evaluated. The electron wave functions can be represented in the form \eqref{ang_mom_addition}, because for alkali atoms the spin-orbit interaction is accurately described by the potential \eqref{V_SO}.

In calculating the matrix elements in the basis of states \eqref{eigen_states_JJz} that are eigenstates for the square of the angular momentum operator and the projection of the angular momentum onto the $z$ axis, it is convenient to use the Wigner-Eckart theorem
\begin{equation}\label{Wigner-Eckart_thm}
\lan n'J'M'|f_{jm}|nJM\ran=\frac1{\sqrt{2J'+1}} C^{J'M'}_{JMjm} \lan n'J'\|f_{j}\|nJ\ran,
\end{equation}
where $f_{jm}$ is an irreducible tensor of rank $j$, $\lan n'J'\|f_{j}\|nJ\ran$ is a reduced matrix element independent of $M'$, $M$, and $m$. Let us obtain the general formulas \cite{LandLifshQM.11} relating the reduced matrix elements for the systems whose wave functions are constructed as in \eqref{ang_mom_addition}. Let the system consist of two subsystems. The bases in the state spaces of these systems are conveniently chosen to be $|n_1,j_1,m_1\ran$ and $|n_2,j_2,m_2\ran$, respectively. Here $n_1$ and $n_2$ are the quantum numbers of the physical quantities that differ from $\mathbf{J}_{1,2}^2$ and $(J_{1,2})_z$ and belong to the complete set of commuting operators defining the basis vectors. We assume that the Hamiltonian of the interaction of systems $(1)$ and $(2)$ commutes with the operators $\mathbf{J}_1^2$, $\mathbf{J}_2^2$ and the operators corresponding to the quantum numbers $n_1$, $n_2$. Let us construct the states of the entire system with definite values of the quantum numbers $n_1$, $n_2$, $j_1$, $j_2$, $J$, and $M$ as
\begin{equation}\label{ang_mom_addition_1}
|n_1,n_2,j_1,j_2,J,M\ran=\sum_{m_1,m_2}C^{JM}_{j_1m_1j_2m_2}|n_1,j_1,m_1\ran\otimes |n_2,j_2,m_2\ran.
\end{equation}
Consider the matrix elements of the operator acting only on the variables of the system $(1)$ and being an irreducible tensor of rank $j$:
\begin{equation}\label{matr_elem_1}
\lan n'_1,n'_2,j'_1,j'_2,J',M'|f^{(1)}_{jm} |n_1,n_2,j_1,j_2,J,M\ran=\frac{1}{\sqrt{2J'+1}} C^{J'M'}_{JMjm} \lan n'_1,n'_2,j'_1,j'_2,J'\|f^{(1)}_{j} \|n_1,n_2,j_1,j_2,J\ran,
\end{equation}
where we have employed the Wigner-Eckart theorem \eqref{Wigner-Eckart_thm}. On the other hand, substituting the state \eqref{ang_mom_addition_1} into this matrix element, we deduce
\begin{equation}\label{matr_elem_2}
\begin{split}
\lan n'_1,n'_2,j'_1,j'_2,J',M'|f^{(1)}_{jm} |n_1,n_2,j_1,j_2,J,M\ran=&\,\frac{\de_{n_2'n_2}\de_{j_2'j_2}}{\sqrt{2j_1'+1}} \sum_{m_1,m_2,m_1'} C^{J'M'}_{j'_1m'_1j_2m_2} C^{JM}_{j_1m_1j_2m_2} C^{j_1'm_1'}_{j_1m_1jm}\times\\
&\times\lan n'_1,j'_1\|f^{(1)}_{j} \|n_1j_1\ran.
\end{split}
\end{equation}
The sum over $m_1$, $m_2$, $m_1'$ is expressed in terms of the $6j$-symbol [(8.7.12) of \cite{Varshalovich}]
\begin{equation}
\sum_{m_1,m_2,m_1'} C^{J'M'}_{j'_1m'_1j_2m_2} C^{JM}_{j_1m_1j_2m_2} C^{j_1'm_1'}_{j_1m_1jm} = (-1)^{j_2+J+j_1'+j} \sqrt{(2J+1)(2j_1'+1)} C^{J'M'}_{JMjm}
\sixj{j_1}{j_2}{J}{J'}{j}{j_1'}.
\end{equation}
As a result, comparing \eqref{matr_elem_2} with \eqref{matr_elem_1}, we arrive at
\begin{equation}\label{fine_struct_reduced}
\begin{split}
\lan n'_1,n'_2,j'_1,j'_2,J'\|f^{(1)}_{j} \|n_1,n_2,j_1,j_2,J\ran =&\,\de_{n_2'n_2}\de_{j_2'j_2}(-1)^{j_2+J+j_1'+j}\times\\
&\times \sqrt{(2J'+1)(2J+1)} \sixj{J'}{J}{j}{j_1}{j_1'}{j_2} \lan n'_1,j'_1\|f^{(1)}_{j} \|n_1j_1\ran,
\end{split}
\end{equation}
where we have used the symmetry properties of the $6j$-symbols.

Let us apply this relation to simplify the matrix elements of the interaction Hamiltonian \eqref{MEjm_IR} describing $Ej$-transitions in the long-wavelength limit. As long as $H_{hfs}$ commutes with $\mathbf{J}^2$ and $\mathbf{I}^2$, the hyperfine structure associated with the total angular momentum of the atom $\mathbf{F}=\mathbf{J}+\mathbf{I}$, where $\mathbf{I}$ is the total angular momentum of the atomic nucleus, is taken into account in accordance with \eqref{fine_struct_reduced} as
\begin{equation}
\lan J',I',F'\|j_j Y_j \|J,I,F\ran=\de_{I'I} (-1)^{I+F+J'+j} \sqrt{(2F'+1)(2F+1)} \sixj{F'}{F}{j}{J}{J'}{I} \lan J'\|j_j Y_j \|J\ran,
\end{equation}
where, for brevity, we do not write out the quantum numbers $n_{1,2}$ and $n'_{1,2}$. The fine structure corresponding to the partition of the total angular momentum of electrons $\mathbf{J}=\mathbf{L}+\mathbf{S}$, where $\mathbf{L}$ is the orbital momentum of electrons and $\mathbf{S}$ is their spin, is described as
\begin{equation}\label{JJ_reduced}
\begin{split}
\lan J'\|j_j Y_j \|J\ran &=\lan L',S',J'\|j_j Y_j \|L,S,J\ran=\\
&=\de_{S'S}(-1)^{S+J+L'+j} \sqrt{(2J'+1)(2J+1)} \sixj{J'}{J}{j}{L}{L'}{S} \lan L'\|j_j Y_j \|L\ran,
\end{split}
\end{equation}
where $S=1/2$ in the case of an alkali atom with one active electron.

To calculate the reduced matrix element on the right-hand side of \eqref{JJ_reduced}, it is sufficient to consider the matrix element
\begin{equation}\label{nLJ_M_C}
\begin{split}
\lan n',L',M'|j_j(k_0r)Y_{jm}(\theta,\vf) |n,L,M\ran &=R^{j;0}_{n'L'J';nLJ}(k_0)\int d\Omega Y^*_{L'M'}Y_{jm}Y_{LM}=\\
&=R^{j;0}_{n'L'J';nLJ}(k_0) \sqrt{\frac{(2j+1)(2L+1)}{4\pi(2L'+1)}} C^{L'0}_{L0j0} C^{L'M'}_{LMjm},
\end{split}
\end{equation}
where we have employed formula (5.9.4) of \cite{Varshalovich} and introduced the notation $R^{j;k}_{n'L'J';nLJ}(k_0)$ for the integral over the radial variable
\begin{equation}\label{matr_elem_radial}
\begin{split}
R^{j;k}_{n'L'J';nLJ}(k_0)&=\int_0^\infty drr^{2+k} R^*_{n'L'J'}(r)j_j(k_0r)R_{nLJ}(r)\approx\frac{k_0^j}{(2j+1)!!} \int_0^\infty drr^{j+k+2} R^*_{n'L'J'}(r)R_{nLJ}(r)=:\\
&=: \frac{k_0^j}{(2j+1)!!} r^{j+k}_{n'L'J';nLJ}.
\end{split}
\end{equation}
The radial wave functions, $R_{nLJ}(r)$, depend on $J$ via the spin-orbit interaction. As for the reduced matrix element, we have
\begin{equation}
\lan L'\|j_j Y_j \|L\ran=R^{j;0}_{n'L'J';nLJ}(k_0) \sqrt{\frac{(2j+1)(2L+1)}{4\pi}} C^{L'0}_{L0j0}.
\end{equation}
The Clebsch-Gordan coefficient in the last formula leads, in particular, to the parity selection rule \eqref{sel_rule_parity} for $Ej$-transitions.

Thus, substituting \eqref{nLJ_M_C} into \eqref{MEjm_IR}, we derive that if we take into account only the fine structure, then, in the long-wavelength limit, the matrix element of the interaction Hamiltonian for $Ej$-transition appearing in \eqref{V_matr_elem} is written as
\begin{equation}\label{reduced_matr_elem}
\begin{split}
\lan n'L'J'M'|H^{app}_{int}[\bs\psi^1_{jm}(k_0)] |nLJM \ran \approx &-ie(-1)^{1/2+J+L'+j}\sqrt{\frac{j+1}{j}} \sqrt{\frac{(2J+1)(2L+1)}{4\pi(2j+1)}}\sixj{J'}{J}{j}{L}{L'}{1/2}\times\\
&\times C^{J'M'}_{JMjm}C^{L'0}_{L0j0} E_{n'L'J'M';nLJM} \frac{k_0^{j-1}}{(2j-1)!!} r^j_{n'L'J';nLJ},
\end{split}
\end{equation}
where $E_{n'L'J'M';nLJM}:=E_{n'L'J'M'}-E_{nLJM}$. The dependence of the state energy on $M$ arises due to the presence of an external magnetic field leading to the Zeeman splitting. Moreover, using Eqs. \eqref{JJ_reduced} and \eqref{nLJ_M_C}, it is not difficult to obtain the matrix element of the dipole moment operator
\begin{equation}\label{dipole_mom_matr_elem}
\begin{split}
\lan n'L'J'M'|(\spe_\la \mathbf{d}) |nLJM \ran=&\,e\sqrt{\frac{4\pi}{3}} \lan n'L'J'M'|rY_{1\la} |nLJM\ran =e(-1)^{1/2+J+L'+1} C^{J'M'}_{JM1\la}\times\\
&\times \sqrt{(2J+1)(2L+1)} \sixj{J'}{J}{1}{L}{L'}{1/2} C^{L'0}_{L010} r_{n'L'J';nLJ}.
\end{split}
\end{equation}
We use this expression to find the dielectric susceptibility of a gas of atoms.

\section{Direct calculation of the multipole matrix elements}\label{App_Mult_Trans_Direct}

In the previous appendix, we saw that the Siegert theorem allows one to express the matrix element of the interaction Hamiltonian through the integral over the radial variable \eqref{reduced_matr_elem} in the long-wavelength limit for $Ej$-transitions. In this appendix we will obtain the explicit expressions for the matrix elements of the interaction Hamiltonian for $Ej$- and $Mj$-transitions without using the long-wavelength approximation.

We need to calculate the matrix element
\begin{equation}
\lan n'L'J'M'|\hat{H}_{int}^{app}[\bs\psi^p_{jm}(k_0,\spx)]|nLJM\ran=(i)+(ii),
\end{equation}
where, as follows from \eqref{H_int_app},
\begin{equation}
\begin{split}
(i)&=-\frac{e}{2m_e} \lan n'L'J'M'|\hat{\mathbf{p}} \bs\psi^p_{jm}(k_0,\spx)
+ \bs\psi^p_{jm}(k_0,\spx) \hat{\mathbf{p}} |nLJM\ran ,\\
(ii)&=\mu_B g_S \lan n'L'J'M'|\boldsymbol{S}\rot(\bs\psi^p_{jm}(k_0,\spx))|nLJM\ran.
\end{split}
\end{equation}
Taking into account only the fine structure, we construct the electron states according to the $LS$-coupling scheme
\begin{equation}
|nLJM\ran=\sum_{M_L,M_S} C^{JM}_{LM_L\frac{1}{2}M_S}R_{nLJ}(r)Y_{LM_L}(\theta,\vf) \chi_{M_S},
\end{equation}
where
\begin{equation}
\chi_{1/2}=
\left[
\begin{array}{c}
1 \\
0 \\
\end{array}
\right],\qquad
\chi_{-1/2}=
\left[
\begin{array}{c}
0 \\
1 \\
\end{array}
\right].
\end{equation}
The generalization of the formulas to account for the hyperfine structure is readily obtained by means of the procedure given in the previous appendix.

Let us start with the matrix element $(ii)$. This matrix element is proportional to
\begin{equation}\label{matr_elem_ii}
\chi_{M'_S}^\dag\mathbf{S}\chi_{M_S}\int_0^\infty  dr r^2\int d\theta
d\vf \sin\theta R^*_{n'L'J'}(r)Y^*_{L'M'_L}(\theta,\vf)\rot\big[\bs\psi^p_{jm}(k_0,\spx)\big]R_{nLJ}(r)Y_{LM_L} (\theta,\vf).
\end{equation}
To calculate this integral we use the curl formula (7.3.55) of \cite{Varshalovich}:
\begin{equation}
\begin{split}
\rot[f(r)\mathbf{Y}^j_{j+1,m}]&=i\sqrt{\frac{j}{2j+1}}(\frac{d}{dr}+\frac{j+2}{r})f(r)\mathbf{Y}^j_{jm},\\
\rot[f(r)\mathbf{Y}^j_{j-1,m}]&=i\sqrt{\frac{j+1}{2j+1}}(\frac{d}{dr}-\frac{j-1}{r})f(r)\mathbf{Y}^j_{jm},\\
\rot[f(r)\mathbf{Y}^j_{jm}]&=i\sqrt{\frac{j}{2j+1}}(\frac{d}{dr}-\frac{j}{r})f(r)\mathbf{Y}^j_{j+1,m}+i\sqrt{\frac{j+1}{2j+1}}(\frac{d}{dr}+\frac{j+1}{r})f(r)\mathbf{Y}^j_{j-1,m}.
\end{split}
\end{equation}
By applying this formula to $\bs\psi^p_{jm}(k_0,\spx)$, the result can be written as
\begin{equation}
\rot\big[\bs\psi^p_{jm}(k_0,\spx) \big]=-ik_0\sum_l \varkappa^p_l j_l(k_0r)\mathbf{Y}^j_{lm},
\end{equation}
where $l=j\pm(1-p)$ and
\begin{equation}
\varkappa^0_{j+1}=\sqrt{\frac{j}{2j+1}},\qquad \varkappa^0_{j-1}=-\sqrt{\frac{j+1}{2j+1}},\qquad
\varkappa^1_{j}=1.
\end{equation}
The integral over the variable $r$ in \eqref{matr_elem_ii} reduces to \eqref{matr_elem_radial}. The angular integrals have the form
\begin{equation}
\int d\theta d\vf \sin\theta Y^*_{L'M'_L}(\theta,\vf)Y_{LM_L}(\theta,\vf)\mathbf{Y}^j_{lm}(\theta,\vf) =
\sqrt{\frac{(2l+1)(2L+1)}{4\pi(2L'+1)}} C^{L'0}_{L 0 l 0}\sum_{a,\s }C^{L' M'_L}_{L M_L l a} C^{j m}_{l a 1 \s } \mathbf{e}_\s,
\end{equation}
where we have used formula \eqref{nLJ_M_C}.

Notice that [(6.2.14) of \cite{Varshalovich}]
\begin{equation}
\chi_{M'_S}^\dag \mathbf{S} \mathbf{e}_\s \chi_{M_S}= \frac{\sqrt{3}}{2} C_{\frac{1}{2} M_S 1 \s}^{\frac{1}{2} M'_S}.
\end{equation}
Then the sums over quantum numbers arising in the matrix element $(ii)$ are written as
\begin{equation}
\begin{split}
& \sum_{M_L,M_S,M'_L,M'_S, \s,a} C^{J'M'}_{L'M'_L\frac{1}{2}M'_S}C^{JM}_{LM_L\frac{1}{2}M_S}
C^{L' M'_L}_{L M_L l a} C^{j m}_{l a 1 \s }  \mathbf{e}_\s \chi_{M'_S}^\dag \mathbf{S}\chi_{M_S} = \\
&= \frac{\sqrt{3}}{2} \sum_{M_L,M_S,M'_L,M'_S, \s,a}  C^{J'M'}_{L'M'_L\frac{1}{2}M'_S}C^{JM}_{LM_L\frac{1}{2}M_S}
C^{L' M'_L}_{L M_L l a} C^{j m}_{l a 1 \s } C_{\frac{1}{2} M_S 1 \s}^{\frac{1}{2} M'_S} =\\
&= \sqrt{\frac{3}{2}}\sum_{\s,a, k , \ka} (-1)^{1+ l - k}\sqrt{(2L'+1)(2J+1)(2k+1)}
C_{l a 1 \s }^{j m} C^{k \ka}_{l, -a; 1, -\s} C^{J'M'}_{JM; k, -\ka} \ninej{J'}{L'}{1/2}{J}{L}{1/2}{k}{l}{1} = \\
&= \sqrt{\frac{3}{2}} \sqrt{(2L'+1)(2J+1)(2j+1)} C^{J'M'}_{JM j m} \ninej{J'}{L'}{1/2}{J}{L}{1/2}{j}{l}{1},
\end{split}
\end{equation}
where formula (8.7.26) of \cite{Varshalovich} has been used. As a result, the contribution $(ii)$ to the matrix element of the interaction Hamiltonian can be cast into the form
\begin{equation}\label{matr_elem_ii_fin}
(ii) = -ik_0 \mu_B g_S  C^{J' M'}_{J M j m} \sum_l  \sqrt{\frac{3(2J+1)(2j+1)(2L+1)(2l+1)}{8\pi}}   C^{L'0}_{L 0 l0}
\ninej{J'}{J}{j}{L'}{L}{l}{1/2}{1/2}{1} \vk^p_l R^{l;0}_{n'L'J',nLJ} .
\end{equation}
The Clebsch-Gordan coefficient preceding the $9j$-symbol ensures, in particular, the parity selection rule \eqref{sel_rule_parity}.

Now we turn to the matrix element $(i)$ where the first momentum operator is supposed to act to the left and the second operator is supposed to act to the right. It is convenient to employ the representation of the mode functions \eqref{multipoles} in the form
\begin{equation}
\bs\psi^p_{jm}(k_0,\spx)=
\left\{
\begin{array}{ll}
j_j(k_0r)\mathbf{Y}^{(0)}_{jm}(\theta,\vf), & \hbox{$p=0$;} \\
(j'_{j}(k_0r) + \frac{j_{j}(k_0r)}{k_0 r})\mathbf{Y}^{(1)}_{jm}(\theta,\vf) +\sqrt{j(j+1)}\frac{j_{j}(k_0r)}{k_0r}
\mathbf{Y}^{(-1)}_{jm}(\theta,\vf), & \hbox{$p=1$,}
\end{array}
\right.
\end{equation}
where
\begin{equation}
\mathbf{Y}^{(1)}_{L M_L}(\theta,\vf) = -i [\mathbf{n}, \mathbf{Y}^{(0)}_{L M_L}(\theta,\vf)],\qquad \mathbf{Y}^{(0)}_{L M_L}(\theta,\vf) = \mathbf{Y}^{L}_{L M_L}(\theta,\vf),\qquad \mathbf{Y}^{(-1)}_{L M_L}(\theta,\vf) = \mathbf{n} Y_{LM_L}(\theta,\vf),
\end{equation}
and $\mathbf{n}$ is a unit vector along the vector $\mathbf{r}$. Substituting the explicit expressions for the electron states and the mode functions $\bs\psi^p_{jm}(k_0,\spx)$ into the matrix element $(i)$, we obtain
\begin{multline}\label{matr_elem_i_1}
(i)=\frac{ i e}{2 m_e}\sum_{M_L,M_S,M'_L} C^{J'M'}_{L'M'_L\frac{1}{2}M_S}C^{JM}_{LM_L\frac{1}{2}M_S}
\int_0^\infty drr^2 \int d\theta  d\vf \sin\theta \bs\psi^p_{jm}(k_0,\spx) \times \\
\times\big[R^*_{n'L'J'}(r)Y^*_{L'M'_L}(\theta,\vf) \nabla(R_{nLJ}(r)Y_{LM_L}(\theta,\vf)) - R_{nLJ}(r)Y_{LM_L}(\theta,\vf)
\nabla(R^*_{n'L'J'}(r)Y^*_{L'M'_L}(\theta,\vf))\big].
\end{multline}
Thus, it is necessary to evaluate the integral over the angular variables and the sums over the magnetic quantum numbers.

Let us consider the magnetic and electric multipole transitions separately. Let us start with the electric transitions and consider the scalar product $ \bs\psi^1_{jm}(k_0,\spx) \nabla(R_{nL}(r)Y_{LM_L}(\theta,\vf))$. In order to proceed, we use the gradient formula \eqref{grad_formula} in the form
\begin{equation}
\nabla(f(r)Y_{LM_L}(\theta,\vf)) = f' \mathbf{Y}^{(-1)}_{L M_L}(\theta,\vf) + \sqrt{L(L+1)} \frac{f}{r}
\mathbf{Y}^{(1)}_{L M_L}(\theta,\vf).
\end{equation}
Besides, we employ the fact that the spherical vectors $\mathbf{Y}^{(1)}_{L M_L}$ and $\mathbf{Y}^{(0)}_{L M_L}$ are orthogonal to $\mathbf{n}$. Then
\begin{equation}
\mathbf{Y}^{(-1)}_{L M_L}(\theta,\vf) \mathbf{Y}^{(1)}_{L M_L}(\theta,\vf)=0,
\end{equation}
and, by using the formula for the inner product of spherical vectors (7.3.100) of \cite{Varshalovich}, we come to
\begin{multline}
\bs\psi^1_{jm}(k_0,\spx) \nabla(R_{nLJ}(r)Y_{LM_L}) = \sqrt{j(j+1)}\frac{j_{j}(k_0r)}{k_0r} R'_{nLJ}(r)Y_{jm}Y_{LM_L}
- \sqrt{L(L+1)} \frac{R_{nLJ}(r)}{r}\times \\
\times(j'_{j}(k_0r) + \frac{j_{j}(k_0r)}{k_0 r}) \sum_{a,\s}(-1)^{j+L+a}
\frac{(2L+1)(2j+1)}{\sqrt{4\pi (2a+1)}} \sixj{j}{j}{1}{L}{L}{a}
C^{a 0}_{j 0 L 0} C^{a \s}_{j m L M_L} Y_{a \s}.
\end{multline}
The $6j$-symbol in this expression has a simple form
\begin{equation}
\sixj{j}{j}{1}{L}{L}{a} = (-1)^{j+L+a+1}\frac{j(j+1) + L(L+1) - a(a+1)}{2\sqrt{j(j+1)(2j+1)L(L+1)(2L+1)}}.
\end{equation}
Therefore, we can write
\begin{multline}
\bs\psi^1_{jm}(k_0,\spx) \nabla(R_{nLJ}(r)Y_{LM_L}) = \sqrt{j(j+1)}\frac{j_{j}(k_0r)}{k_0r} R'_{nLJ}(r)Y_{jm}Y_{LM_L}
+\frac{R_{nLJ}(r)}{r} (j'_{j}(k_0r) + \frac{j_{j}(k_0r)}{k_0 r})  \times \\
\times\sum_{a,\s} (j(j+1) + L(L+1) - a(a+1))\frac{\sqrt{(2L+1)(2j+1)}}{2\sqrt{4\pi j(j+1) (2a+1)}}
C^{a 0}_{j 0 L 0} C^{a \s}_{j m L M_L} Y_{a \s}.
\end{multline}
Similarly, using the property (3.5.11) of \cite{Varshalovich},
\begin{equation}
Y^*_{L'M'_L}(\theta,\vf) = (-1)^{M'_L} Y_{L',-M'_L}(\theta,\vf),
\end{equation}
we obtain for the second term in \eqref{matr_elem_i_1}:
\begin{multline}
\bs\psi^1_{jm}(k_0,\spx) \nabla(R^*_{n'L'J'}(r)Y^*_{L'M'_L}) = \sqrt{j(j+1)}\frac{j_{j}(k_0r)}{k_0r}
R'^*_{n'L'J'}(r)Y_{jm}Y^*_{L'M'_L} -  \sqrt{L'(L'+1)} \frac{R^*_{n'L'J'}(r)}{r}\times \\
\times(j'_{j}(k_0r) + \frac{j_{j}(k_0r)}{k_0 r}) \sum_{a,\s} (-1)^{M'_L + \s}
\frac{(2L'+1)(2j+1)}{\sqrt{4\pi (2a+1)}} \sixj{j}{j}{1}{L'}{L'}{a} C^{a 0}_{j 0 L' 0} C^{a, -\s}_{j m; L', -M'_L} Y^*_{a \s}.
\end{multline}
Substituting the explicit expression for the $6j$-symbol, we have
\begin{multline}
\bs\psi^1_{jm}(k_0,\spx) \nabla(R^*_{n'L'J'}(r)Y^*_{L'M'_L}) = \sqrt{j(j+1)}\frac{j_{j}(k_0r)}{k_0r} R'^*_{n'L'J'}(r)
Y_{jm}Y^*_{L'M'_L}  +  \frac{R^*_{n'L'J'}(r)}{r} (j'_{j}(k_0r) + \frac{j_{j}(k_0r)}{k_0 r})\times \\
\times \sum_{a,\s} (j(j+1) + L'(L'+1) - a(a+1)) \frac{\sqrt{(2a+1)(2j+1) }}{2\sqrt{4\pi j(j+1)(2L'+1)}}
C^{L' 0}_{j 0 a 0} C^{L' M'_L }_{j m a \s} Y^*_{a \s}.
\end{multline}
On substituting this expression into \eqref{matr_elem_i_1}, the integrals over the angular variables are easily calculated using the relation \eqref{nLJ_M_C} and the orthonormality property of spherical functions. As a result, we have for $Ej$-transitions
\begin{multline}\label{currentInts}
(i)  =
\frac{ i e}{2 m_e} (-1)^{1/2 + J + L' + j}  C^{J' M' }_{J M j m } C^{L' 0}_{L 0 j 0 } \sixj{J'}{J}{j}{L}{L'}{1/2}  \sqrt{\frac{(2J+1)(2L+1)(2j+1)}{4\pi }} \sqrt{\frac{j+1}{j}} \times \\
\times \bigg[
\frac{j}{k_0} \int_0^\infty dr r j_{j}(k_0r) W(r) + \frac{L(L+1) - L'(L'+1)}{j+1}  \big[ (\vk^0_{j-1})^2 R^{j-1;-1}_{n'L'J';nLJ} - (\vk^0_{j+1})^2 R^{j+1;-1}_{n'L'J';nLJ}\big]\bigg],
\end{multline}
where $W(r) = R'_{nLJ}(r) R^*_{n'L'J'}(r) - R_{nLJ}(r) R'^*_{n'L'J'}(r) $ is the Wronskian of solutions of the radial Schr\"{o}dinger equation for the Hamiltonian \eqref{Hamilt_electron}. The matrix element of the interaction Hamiltonian for $Ej$-transitions is given by the sum of the matrix elements \eqref{matr_elem_ii_fin} and \eqref{currentInts}.

Now we consider the matrix element $(i)$ for magnetic transitions. Carrying out calculations similar to those done above, we obtain for the inner product
\begin{multline}\label{scalar_prod_mag}
\bs\psi^0_{jm}(k_0,\spx) \nabla(R_{nLJ}(r)Y_{LM_L}) = i j_j(k_0r) \sqrt{L(L+1)}\frac{R_{nLJ}(r)}{r}
(\mathbf{n}, \mathbf{Y}^j_{j m}, \mathbf{Y}^L_{LM_L}) = \\
= j_j(k_0r) \frac{R_{nLJ}(r)}{r}(2j+1) (2L+1)\sqrt{\frac{3L(L+1)}{2\pi}} \sum_{a,\s,\xi=\pm 1}
\vk^0_{a-\xi} C^{a+\xi, 0}_{ L 0 j 0}C^{a  \s}_{ L M_L j m} \ninej{j}{j}{1}{L}{L}{1}{a}{a+\xi}{1} Y_{a \s},
\end{multline}
where formula (7.10.101) of \cite{Varshalovich} for the vector product of spherical vectors and the relations (7.8.71) and (7.8.72) of \cite{Varshalovich} for the inner product of the vector $\mathbf{n}$ with spherical vectors have been used. We have also denoted the triple product of vectors as $(\mathbf{a},\mathbf{b},\mathbf{c})$. Having substituted \eqref{scalar_prod_mag} into the matrix element \eqref{matr_elem_i_1}, it is clear that the angular integrals are readily evaluated. In addition, the $9j$-symbol entering into \eqref{scalar_prod_mag} can be rewritten in terms of the $6j$-symbol [(10.9.6) of \cite{Varshalovich}]. Then the matrix element $(i)$ for $Mj$-transitions is given by
\begin{equation}
\begin{split}
(i)  =&  -\frac{i e}{m_e} (-1)^{1/2 + J + L' + j}  C^{J' M' }_{J M j m } \sixj{J'}{J}{j}{L}{L'}{1/2}
(2j+1) \sqrt{\frac{(2J+1)(2L+1)}{4\pi}} \times \\
&\times R^{j;-1}_{n'L'J';nLJ}  \sum_{\xi=\pm1}  \big(L' + \frac{1+\xi}{2}\big)^{3/2}
C^{L'+\xi, 0}_{ L 0 j 0}\sixj{j}{j}{1}{L'+\xi}{L'}{L}.
\end{split}
\end{equation}
Taking into account that
\begin{equation}
\sum_{\xi=\pm1}  \big(L' + \frac{1+\xi}{2}\big)^{3/2}
C^{L'+\xi, 0}_{ L 0 j 0}\sixj{j}{j}{1}{L'+\xi}{L'}{L}=(2L'+1)'\sqrt{L'} C^{L'-1, 0}_{ L 0 j 0}\sixj{j}{j}{1}{L'-1}{L'}{L},
\end{equation}
we eventually arrive at
\begin{equation}\label{matr_elem_i_Mj_fin}
\begin{split}
(i)  =&  -\frac{i e}{m_e} (-1)^{1/2 + J + L' + j}  C^{J' M' }_{J M j m } C^{L'-1, 0}_{ L 0 j 0} \sixj{J'}{J}{j}{L}{L'}{1/2}
\sixj{j}{j}{1}{L'-1}{L'}{L} \times \\
&\times (2L'+1)(2j+1) \sqrt{\frac{L'(2J+1)(2L+1)}{4\pi}} R^{j;-1}_{n'L'J';nLJ}.
\end{split}
\end{equation}
The Clebsch-Gordan coefficient standing in front of the $6j$-symbols ensures, in particular, the fulfillment of the parity selection rule \eqref{sel_rule_parity}. The matrix element of the interaction Hamiltonian for $Mj$-transitions is given by the sum of the matrix elements \eqref{matr_elem_ii_fin} and \eqref{matr_elem_i_Mj_fin}.

Let us consider the long-wavelength limit when $k_0 r_B \ll 1$. The radial integrals in this limit are considerably simplified (see \eqref{matr_elem_radial}). Then, in the leading order in $k_0r_B$, we obtain for $Mj$-transitions
\begin{equation}\label{Mj_longwavelength_i}
\begin{split}
(i)\approx &\frac{i e}{m_e} (-1)^{1/2 + J + L}  C^{J' M' }_{J M j m }C^{L'-1, 0}_{ L 0 j 0} \sixj{J'}{J}{j}{L}{L'}{1/2} \sixj{j}{j}{1}{L'-1}{L'}{L}
\times \\
&\times (2L'+1)\sqrt{\frac{L'(2J+1)(2L+1)}{4\pi}} \frac{k_0^j r^{j-1}_{n'L'J';nLJ}}{(2j-1)!!}.
\end{split}
\end{equation}
The matrix element \eqref{matr_elem_ii_fin} is reduced to
\begin{equation}\label{Mj_longwavelength_ii}
(ii)\approx i \mu_B g_S  C^{J' M'}_{J M j m}  \sqrt{\frac{3(2J+1)(2L+1)(j+1)(2j-1)}{8\pi}}   C^{L'0}_{L 0;j-1,0}
\ninej{J'}{J}{j}{L'}{L}{j-1}{1/2}{1/2}{1} \frac{k_0^j r^{j-1}_{n'L'J';nLJ}}{(2j-1)!!} .
\end{equation}
As we see, the contributions $(i)$ and $(ii)$ are of the same order of smallness.

As for $Ej$-transitions in the long-wavelength limit, we consider at first the matrix element $(i)$. To simplify the integrals in \eqref{currentInts}, we rewrite the Wronskian using the radial Schr\"{o}dinger equation as
\begin{equation}
(r^2 W)' = r^2 R_{nLJ}R^*_{n'L'J'}\Big[ \frac{L(L+1) - L'(L'+1)}{ r^2} + 2\mu(V_{nL}(r) - V_{n'L'}(r) )
- \frac{\al \mu}{m^2_e r^3}((\mathbf{L'}- \mathbf{L})\mathbf{S}) + 2\mu E_{n'L'J'; nLJ}\Big].
\end{equation}
Then, expanding the spherical Bessel function in the neighborhood of zero \eqref{Bessel_expns}, replacing $\mu\rightarrow m_e$, and integrating by parts in the first term in square brackets in \eqref{currentInts},  we obtain the approximate expression for the expression in square brackets in \eqref{currentInts}:
\begin{equation}
-2 m_e E_{n'L'J'; nLJ} \frac{k_0^{j-1}}{(2j+1)!!}  r^{j}_{n'L'J';nLJ},
\end{equation}
where we have discarded the contribution of the spin-orbit interaction and neglected the dependence of the potential on the quantum numbers \cite{Marinescu1994,Sibalic2017ARC}. Under these approximations, we obtain for $Ej$-transitions in the long-wavelength limit
\begin{equation}
\begin{split}
(i)  \approx&  -i e(-1)^{1/2 + J + L' + j} C^{L' 0}_{L 0 j 0 } C^{J' M' }_{J M j m }
\sqrt{\frac{(2J+1)(2L+1)(2j+1)}{4\pi }}
\sixj{J'}{J}{j}{L}{L'}{1/2}\times\\
&\times \sqrt{\frac{j+1}{j}} E_{n'L'J'; nLJ} \frac{k_0^{j-1}}{(2j+1)!!}  r^{j}_{n'L'J';nLJ}.
\end{split}
\end{equation}
This expression is exactly the same as that follows from the Siegert theorem \eqref{reduced_matr_elem}. Employing the expansion \eqref{matr_elem_radial} for the radial integrals in the long-wavelength limit we see for $Ej$-transitions that the matrix element \eqref{matr_elem_ii_fin}, which also contributes to the matrix element of the interaction Hamiltonian, is by the order of magnitude
\begin{equation}
(ii)/(i) \sim k_0^2/(m_e E_{n'L'J'; nLJ}),
\end{equation}
i.e., it is negligibly small as compared with the matrix element $(i)$ for $k_0\approx |E_{n'L'J'; nLJ}|$ and $k_0\ll m_e$.


\begin{thebibliography}{999}



\bibitem{Sedlacek2012}
J.~A. Sedlacek, A. Schwettmann, H. K\"{u}bler, R. L\"{o}w, T. Pfau, J.~P. Shaffer,
Microwave electrometry with Rydberg atoms in a vapour cell using bright atomic resonances,
Nature Phys. \textbf{8}, 819 (2012).

\bibitem{Sedlacek2013}
J.~A. Sedlacek, A. Schwettmann, H. K\"{u}bler, J.~P. Shaffer,
Atom-based vector microwave electrometry using Rubidium Rydberg atoms in a vapor cell,
Phys. Rev. Lett. \textbf{111}, 063001 (2013).

\bibitem{Anderson2014}
D.~A. Anderson, A. Schwarzkopf, S.~A. Miller, N. Thaicharoen, G. Raithel,
Two-photon microwave transitions and strong-field effects in a room-temperature Rydberg-atom gas,
Phys. Rev. A \textbf{90}, 043419 (2014).

\bibitem{Holloway2017}
C.~L. Holloway, M.~T. Simons, J.~A. Gordon, A. Dienstfrey, D.~A. Anderson, G. Raithel,
Electric field metrology for SI traceability: Systematic measurement uncertainties in electromagnetically induced transparency in atomic vapor,
J. Appl. Phys. \textbf{121}, 233106 (2017).

\bibitem{Simons2019}
M.~T. Simons, A.~H. Haddab, J.~A. Gordon, C.~L. Holloway,
A Rydberg atom-based mixer: Measuring the phase of a radio frequency wave,
Appl. Phys. Lett. \textbf{114}, 114101 (2019).

\bibitem{Gordon2019}
J.~A. Gordon, M.~T. Simons, A.~H. Haddab, C.~L. Holloway,
Weak electric-field detection with sub-$1$ Hz resolution at radio frequencies using a Rydberg atom-based mixer,
AIP Advances \textbf{9}, 045030 (2019).

\bibitem{Song2019}
Z. Song, H. Liu, X. Liu, W. Zhang, H. Zou, J. Zhang, J. Qu,
Rydberg-atom-based digital communication using a continuously tunable radio-frequency carrier,
Optics Express \textbf{27}, 8848 (2019).

\bibitem{Jau2020}
Y.-Y. Jau, T. Carter,
Vapor-cell-based atomic electrometry for detection frequencies below $1$ kHz,
Phys. Rev. Applied \textbf{13}, 054034 (2020).

\bibitem{Stelmashenko2020}
E.~F. Stelmashenko, O.~A. Klezovich, V.~N. Baryshev, V.~A. Tishchenko, I.~Yu. Blinov, V.~G. Palchikov, V.~D. Ovsyannikov,
Measuring the electric field strength of microwave radiation at the frequency of the radiation transition between Rydberg states of atoms ${}^{85}$Rb,
Opt. Spectrosc. \textbf{128}, 1067 (2020).

\bibitem{Xue2021}
Y. Xue, Y. Jiao, L. Hao, J. Zhao,
Microwave two-photon spectroscopy of cesium Rydberg atoms,
Optics Express \textbf{29}, 43827 (2021).

\bibitem{Anderson2021}
D.~A. Anderson, R.~E. Sapiro, G. Raithel,
A self-calibrated SI-traceable Rydberg atom-based radio frequency electric field probe and measurement instrument,
IEEE Transactions on Antennas and Propagation \textbf{69}, 5931 (2021).


\bibitem{Robinson2021}
A.~K. Robinson, N. Prajapati, D. Senic, M.~T. Simons, C.~L. Holloway,
Determining the angle-of-arrival of a radio-frequency source with a Rydberg atom-based sensor,
Appl. Phys. Lett. \textbf{118}, 114001 (2021).

\bibitem{Anderson2021AMFM}
D.~A. Anderson, R.~E. Sapiro, G. Raithel,
An atomic receiver for AM and FM radio communication,
IEEE Transactions on Antennas and Propagation \textbf{69}, 2455 (2021).

\bibitem{Liu2022}
X.-H. Liu \textit{et al}.,
Continuous-frequency microwave heterodyne detection in an atomic vapor cell,
Phys. Rev. Applied \textbf{18}, 054003 (2022).

\bibitem{Yuan2023}
J. Yuan, T. Jin, L. Xiao, S. Jia, L. Wang,
A Rydberg atom-based receiver with amplitude modulation technique for the fifth-generation millimeter-wave wireless communication,
IEEE Antennas and Wireless Propagation Letters \textbf{22}, 2580 (2023).

\bibitem{Ryabtsev2024}
I.~I. Ryabtsev, V.~M. Entin, D.~B. Tretyakov, E.~A. Yakshina, I.~I. Beterov, Yu.~Ya. Pechersky,
Quantum sensors of electric fields based on highly excited Rydberg atoms,
Radiophys. Quantum Electronics \textbf{67}, 1 (2024).

\bibitem{Hao2024}
J. Hao \textit{et al}.,
Microwave electrometry with Rydberg atoms in a vapor cell using microwave amplitude modulation,
Chin. Phys. B \textbf{33}, 050702 (2024).

\bibitem{Nowosielski2024}
J. Nowosielski \textit{et al}.,
Warm Rydberg atom-based quadrature amplitude-modulated receiver,
Optics Express \textbf{32}, 30027 (2024).


\bibitem{Gong2024}
T. Gong \textit{et al}.,
Rydberg atomic quantum receivers for classical wireless communication and sensing,
arXiv:2409.14501.

\bibitem{Chen2025}
Y. Chen \textit{et al}.,
Harnessing Rydberg atomic receivers: From quantum physics to wireless communications,
arXiv:2501.11842.

\bibitem{Thide07}
B. Thid\'{e} \textit{et al}.,
Utilization of photon orbital angular momentum in the low-frequency radio domain,
Phys. Rev. Lett. \textbf{99}, 087701 (2007).

\bibitem{Mohammadi10}
S.~M. Mohammadi \textit{et al}.,
Orbital angular momentum in radio -- A system study,
IEEE Transactions on Antennas and Propagation \textbf{58}, 565 (2010).

\bibitem{Roadmap16}
H. Rubinsztein-Dunlop \textit{et al}.,
Roadmap on structured light,
J. Opt. \textbf{19}, 013001 (2017).

\bibitem{SerboNew}
B.~A. Knyazev, V.~G. Serbo,
Beams of photons with nonzero projections of orbital angular momenta: New results,
Phys. Usp. \textbf{61}, 449 (2018).

\bibitem{ZWYB20}
K. Zhang, Y. Wang, Y. Yuan, S.~N. Burokur,
A review of orbital angular momentum vortex beams generation: From traditional methods to metasurfaces,
Appl. Sci. \textbf{10}, 1015 (2020).

\bibitem{OAMPM}
R. Chen, H. Zhou, M. Moretti, X. Wang, J. Li,
Orbital angular momentum waves: Generation, detection and emerging applications,
IEEE Communications Surveys \& Tutorials \textbf{22}, 840 (2020).

\bibitem{Noor22}
S.~K. Noor \textit{et al}.,
A review of orbital angular momentum vortex waves for the next generation wireless communications,
IEEE Access \textbf{10}, 89465 (2022).

\bibitem{JiangWerner22}
Z.~H. Jiang, D.~H. Werner (Eqs.),
\textsl{Electromagnetic Vortices: Wave Phenomena and Engineering Applications}
(Wiley, Hoboken, 2022).



\bibitem{ADKL20}
V.~P. Aksenov, V.~V. Dudorov, V.~V. Kolosov, M.~E. Levitsky,
Synthesized vortex beams in the turbulent atmosphere,
Frontiers in Physics \textbf{8}, 143 (2020).

\bibitem{Khan22}
M.~I.~W. Khan \textit{et al}.,
A 0.31-THz orbital-angular-momentum (OAM) wave transceiver in CMOS with bits-to-OAM mode mapping,
IEEE Journal of Solid-State Circuits \textbf{57}, 1344 (2022).

\bibitem{Willner22}
A.~E. Willner \textit{et al}.,
High capacity terahertz communication systems based on multiple orbital-angular-momentum beams,
J. Opt. \textbf{24}, 124002 (2022).


\bibitem{Li20}
Z. Li \textit{et al}.,
The limits of effective degrees of freedom in UCA based orbital angular momentum multiplexed communications,
Sci. Rep. \textbf{10}, 5216 (2020).

\bibitem{CZLZ23}
R. Chen, J. Zhou, W.-X. Long, W. Zhang,
Hybrid circular array and Luneberg lens for long-distance OAM wireless communications,
IEEE Transactions on Communications \textbf{71}, 485 (2023).


\bibitem{Afanas13}
A. Afanasev, C.~E. Carlson, A. Mukherjee,
Off-axis excitation of hydrogen-like atoms by twisted photons,
Phys. Rev. A \textbf{88}, 033841 (2013).


\bibitem{Mukherjee2018}
K. Mukherjee, S. Majumder, P.~K. Mondal, B. Deb,
Interaction of a Laguerre–Gaussian beam with trapped Rydberg atoms,
J. Phys. B: At. Mol. Opt. Phys. \textbf{51}, 015004 (2018).

\bibitem{Duan2019}
Y. Duan, R.~A. M\"{u}ller, A. Surzhykov,
Selection rules for atomic excitation by twisted light,
J. Phys. B: At. Mol. Opt. Phys. \textbf{52}, 184002 (2019).


\bibitem{Lange2022}
R. Lange, N. Huntemann, A.~A. Peshkov, A. Surzhykov, E. Peik,
Excitation of an electric octupole transition by twisted light,
Phys. Rev. Lett. \textbf{129}, 253901 (2022).

\bibitem{Peshkov2023}
A.~A. Peshkov, Y.~M. Bidasyuk, R. Lange, N. Huntemann, E. Peik, A. Surzhykov,
Interaction of twisted light with a trapped atom: Interplay between electronic and motional degrees of freedom,
Phys. Rev. A \textbf{107}, 023106 (2023).

\bibitem{KazSokExcit}
P.~O. Kazinski, A.~A. Sokolov,
Excitation of multipolar transitions in nuclei by twisted photons.
Phys. Atom. Nucl. \textbf{87}, 561 (2024).

\bibitem{GiantResonance}
Z.-W. Lu \textit{et al}.,
Manipulation of giant multipole resonances via vortex $\ga$ photons,
Phys. Rev. Lett. \textbf{131}, 202502 (2023).

\bibitem{Kirschbaum2024}
T. Kirschbaum, T. Schumm, A. P\'{a}lffy,
Photoexcitation of the ${}^{229}$Th nuclear clock transition using twisted light,
Phys. Rev. C \textbf{110}, 064326 (2024).

\bibitem{Wang2024}
Z. Wang, C. Zhang,
Rydberg atoms enabled miniaturization of quantum state OAM on-off keying system,
2024 IEEE International Conference on Communications Workshops (ICC Workshops), Denver, CO, USA, 2024, pp. 761-766.


\bibitem{Gea-Banacloche1995}
J. Gea-Banacloche, Y. Li, S. Jin, M. Xiao,
Electromagnetically induced transparency in ladder-type inhomogeneously broadened media: Theory and experiment,
Phys. Rev. A \textbf{51}, 576 (1995).


\bibitem{Sandhya1997}
S.~N. Sandhya, K.~K. Sharma,
Atomic coherence effects in four-level systems: Doppler-free absorption within an electromagnetically-induced-transparency window,
Phys. Rev. A \textbf{55}, 2155 (1997).

\bibitem{McGloin2001}
D. McGloin, D.~J. Fulton, M.~H. Dunn,
Electromagnetically induced transparency in $N$-level cascade schemes,
Optics Commun. \textbf{190}, 221 (2001).

\bibitem{Cohen-Tannoudji_book_2004}
C. Cohen-Tannoudji, J. Dupont-Roc, G. Grynberg,
\textsl{Atom-Photon Interactions: Basic Processes and Applications}
(Wiley-VCH Verlag, Weinheim, 2004).

\bibitem{Fleischhauer2005}
M. Fleischhauer, A. Imamolu, J.~P. Marangos,
Electromagnetically induced transparency: Optics in coherent media,
Rev. Mod. Phys. \textbf{77}, 633 (2005).


\bibitem{Marinescu1994}
M. Marinescu, H.~R. Sadeghpour, A. Dalgarno,
Dispersion coeffcients for alkali-metal dimers,
Phys. Rev. A \textbf{49}, 982 (1994).

\bibitem{Sibalic2017ARC}
N. \v{S}ibalic, J.~D. Pritchard, C.~S. Adams, K.~J. Weatherill,
ARC: An open-source library for calculating properties of alkali Rydberg atoms,
Computer Physics Communications \textbf{220}, 319 (2017).


\bibitem{Frish1963}
S.~E. Frish,
\textsl{Optical Spectra of Atoms}
(Fizmatlit, Moscow, 1963)
[in Russian].

\bibitem{LandLifshQM.11}
L.~D. Landau, E.~M. Lifshitz,
\textsl{Quantum Mechanics. Non-relativistic Theory}
(Pergamon, Oxford, 1991).

\bibitem{Steck2024Cs}
D.~A. Steck,
Cesium D line data,
http://steck.us/alkalidata.

\bibitem{Steck_book_2025}
D.~A. Steck,
Quantum and Atom Optics,
http://steck.us/teaching.

\bibitem{Ripka2022}
F. Ripka, C. Liu, M. Schmidt, H. K\"{u}bler, J.~P. Shaffer,
Rydberg atom-based radio frequency electrometry: Hyperfine effects,
Proceedings Volume 12016, Optical and Quantum Sensing and Precision Metrology II, San Francisco, CA, USA, 2022, pp. 102-107.

\bibitem{PDG2024}
S. Navas \textit{et al}. (Particle Data Group),
Review of particle physics,
Phys. Rev. D \textbf{110}, 030001 (2024).

\bibitem{Steck2024Rb}
D.~A. Steck,
Rubidium 85 D Line Data,
http://steck.us/alkalidata.

\bibitem{BjoDre}
J.~D. Bjorken, S.~D. Drell,
\textsl{Relativistic Quantum Theory} Vol. I: \textsl{Relativistic Quantum Mechanics}
(McGraw-Hill, New York, 1964).

\bibitem{KazLaz2021}
P.~O. Kazinski, G.~Yu. Lazarenko,
Transition radiation from a Dirac particle wave packet traversing a mirror,
Phys. Rev. A \textbf{103}, 012216 (2021).

\bibitem{KazMokRyk2025}
P.~O. Kazinski, M.~V. Mokrinskiy, V.~A. Ryakin,
Surface photoelectric effect by twisted photons as a source of twisted electrons,
Proc. R. Soc. A \textbf{481}, 20240777 (2025).

\bibitem{Happer2010}
W. Happer, Y.-Y. Jau, T. Walker,
\textsl{Optically Pumped Atoms}
(Wiley-VCH Verlag, Weinheim, 2010).


\bibitem{Downes2023}
L. Downes,
Simple Python tools for modelling few-level atom-light interactions,
J. Phys. B: At. Mol. Opt. Phys. \textbf{56}, 223001 (2023).

\bibitem{LandLifQED}
V.~B. Berestetskii, E.~M. Lifshitz, L.~P. Pitaevskii,
\textsl{Quantum Electrodynamics}
(Butterworth-Heinemann, Oxford, 1982).

\bibitem{Varshalovich}
D.~A. Varshalovich, A.~N. Moskalev, V.~K. Khersonskii,
\textsl{Quantum Theory of Angular Momentum}
(World Scientific, Singapore, 1988).

\bibitem{Prud2}
A.~P. Prudnikov, Y.~A. Brychkov, O.~I. Marichev,
\textsl{Integrals and Series: Special Functions}
(Gordon and Breach Sci. Publ., Amsterdam, 1998), vol. 2.



\bibitem{Pandey2013}
K. Pandey,
Role of different types of subsystems in a doubly driven $\Lambda$ system in ${}^{87}$Rb,
Phys. Rev. A \textbf{87}, 043838 (2013).



\bibitem{Elgee2023}
P.~K. Elgee, J.~C. Hill, K.-J.~E. LeBlanc, G.~D. Ko, P.~D. Kunz, D.~H. Meyer, K.~C. Cox,
Satellite radio detection via dual-microwave Rydberg spectroscopy,
Appl. Phys. Lett. \textbf{123}, 084001 (2023).

\bibitem{Allinson2024}
G. Allinson, M.~J. Jamieson, A.~R. Mackellar, L. Downes, C.~S. Adams, K.~J. Weatherill,
Simultaneous multiband radio-frequency detection using high-orbital-angular-momentum states in a Rydberg-atom receiver,
Phys. Rev. Research \textbf{6}, 023317 (2024).

\bibitem{Wen2024}
W. Wen \textit{et al}.,
Rydberg-atom-based multiband frequency-hopping communication receiver using five-level atomic system,
Optics Express \textbf{32}, 42872 (2024).

\bibitem{Ryabtsev2011}
I.~I. Ryabtsev, I.~I. Beterov, D.~B. Tretyakov, V.~M. Entin, E.~A. Yakshina,
Doppler- and recoil-free laser excitation of Rydberg states via three-photon transitions,
Phys. Rev. A \textbf{84}, 053409 (2011).

\bibitem{Fulton1995}
D.~J. Fulton, S. Shepherd, R.~R. Moseley, B.~D. Sinclair, M.~H. Dunn,
Continuous-wave electromagnetically induced transparency: A comparison of $V$, $\Lambda$, and cascade systems,
Phys. Rev. A \textbf{52}, 2302 (1995).


\bibitem{Gong2025}
T. Gong, C. Yuen, C.~M.~S. See, M. Debbah, L. Hanzo,
Rydberg atomic quantum receivers for the multi-user MIMO uplink,
arXiv:2501.18382.

\bibitem{Richardson2025}
D. Richardson, J. Dee, B.~N. Kayim, B.~C. Sawyer, R. Wyllie, R.~T. Lee, R.~S. Westafer,
Study of angle of arrival estimation with linear arrays of simulated Rydberg atom receivers,
APL Quantum \textbf{2}, 016123 (2025).

\bibitem{Cui2025}
M. Cui, Q. Zeng, K. Huang,
Towards atomic MIMO receivers,
IEEE Journal on Selected Areas in Communications \textbf{43}, 659 (2025).

\bibitem{Kim2025}
H. Kim, H. Park, S. Kim,
Quantum-MUSIC: Multiple signal classification for quantum wireless sensing,
IEEE Wireless Communications Letters \textbf{14}, 1623 (2025).


\bibitem{annalsOAMmult}
P.~O. Kazinski, P.~S. Korolev, G.~Yu. Lazarenko, V.~A. Ryakin,
Multiplexing signals with twisted photons by a circular arc phased array,
Annals Phys. \textbf{462}, 169610 (2024).


\bibitem{Zheng2015}
S. Zheng \textit{et al}.,
Orbital angular momentum mode-demultiplexing scheme with partial angular receiving aperture,
Optics Express \textbf{23}, 12251 (2015).

\bibitem{Zheng2022}
S. Zheng \textit{et al}.,
Plane spiral OAM mode-group orthogonal multiplexing communication using partial arc sampling receiving scheme,
IEEE Transactions on Antennas and Propagation \textbf{70}, 1352 (2022).

\bibitem{Chen2022}
X. Chen, W. Xue,
OAM Communications in multipath environments,
in \textsl{Electromagnetic Vortices: Wave Phenomena and Engineering Applications}
edited by Z.~H. Jiang, D.~H. Werner
(Wiley, Hoboken, 2022).

\bibitem{YaSaLee22}
Y. Yagi, H. Sasaki, D. Lee,
Prototyping of 40 GHz band orbital angular momentum multiplexing system and evaluation of field wireless transmission experiments,
IEEE Access \textbf{10}, 130040 (2022).


\bibitem{Giat2025}
A. Giat, K. Levi, O. Nefesh, L. Stern,
Subwavelength micromachined vapor-cell based Rydberg sensing,
arXiv:2504.09559.

\bibitem{Meyer2021}
D.~H. Meyer, P.~D. Kunz, K.~C. Cox,
Waveguide-coupled Rydberg spectrum analyzer from $0$ to $20$ GHz,
Phys. Rev. Applied \textbf{15}, 014053 (2021).

\bibitem{Amarloo2025}
H. Amarloo, M. Noaman, S.-P. Yu , D. Booth, S. Mirzaee, R. Pandiyan, F. Christaller, J.~P. Shaffer,
A photonic crystal receiver for Rydberg atom-based sensing,
Commun. Eng. \textbf{4}, 70 (2025).

\bibitem{AhiezSit}
A.~I. Akhiezer, A.~G. Sitenko, V.~K. Tartakovskii,
\textsl{Nuclear Electrodynamics}
(Springer, Heidelberg, 1994).



\end{thebibliography}
\end{document}